\documentclass[11pt]{article}
\usepackage{caption}
\usepackage{scalerel}
\usepackage{amsthm,latexsym,amsxtra,graphicx,appendix,epstopdf,feynmf,hyperref,setspace,fix-cm}
\usepackage[normalem]{ulem}
\usepackage{makeidx}
\usepackage{fullpage}
\usepackage{amsmath}
\usepackage{amssymb}
\usepackage{setspace}
\usepackage{bbm}
\usepackage{dsfont}
\usepackage{epsfig}
\usepackage[font=footnotesize,labelfont=bf,justification=centerlast,width=.94\textwidth]{caption}
\usepackage{cite}
 \usepackage{multirow}
\usepackage{array,booktabs}
\usepackage{subfigure}
\usepackage{mathtools}
\usepackage{color}
\usepackage[makeroom]{cancel}
\usepackage{tikz}
\usepackage{pgfplots}
\usetikzlibrary{intersections, pgfplots.fillbetween}
\usepackage{tensor}
\usepackage{simplewick}
\usepackage{frcursive}

\usepackage{pgfornament,multicol}

\usepackage{caption}
\usepackage{hyperref}

\usepackage[skip=0.2cm, indent=0pt]{parskip}

\hypersetup{
    bookmarks=true,%
    colorlinks,%
    citecolor=blue,%
    filecolor=blue,%
    linkcolor=blue,%
    urlcolor=blue
}
\usepackage{float}
\usepackage{slashed}
\usepackage{tikz}
\usetikzlibrary{decorations.pathmorphing}

\newcommand{\bi}{\begin{itemize}}
\newcommand{\ei}{\end{itemize}}
\newcommand{\bea}{\begin{eqnarray}}
\newcommand{\eea}{\end{eqnarray}}
\newcommand{\be}{\begin{equation}}
\newcommand{\ee}{\end{equation}}

\def\frakr{\mathfrak{r}}

\def\XXint#1#2#3{{\setbox0=\hbox{$#1{#2#3}{\int}$}
     \vcenter{\hbox{$#2#3$}}\kern-.5\wd0}}

\def\={\, = \,}

\numberwithin{equation}{section}
\allowdisplaybreaks  

\begin{document}

\vspace*{2.5cm}
\begin{center}
{ \Large \textsc{Cosmological Observatories}} \\ \vspace*{2.3cm}

\end{center}

\begin{center}
Dionysios Anninos$^{1,2}$~, Dami\'an A. Galante$^{1}$~, and Chawakorn Maneerat$^{1}$ 
\end{center}
\begin{center}
{
\footnotesize
\vspace{0.4cm}
$^1$Department of Mathematics, King's College London, Strand, London WC2R 2LS, UK
\\
{$^2$ Instituut voor Theoretische Fysica, KU Leuven, Celestijnenlaan 200D, B-3001 Leuven, Belgium} 
}
\end{center}
\begin{center}
{\textsf{\footnotesize{
dionysios.anninos@kcl.ac.uk, damian.galante@kcl.ac.uk, chawakorn.maneerat@kcl.ac.uk}} } 
\end{center}

\vspace*{0.5cm}

\vspace*{1.5cm}
\begin{abstract}
\noindent
We study the static patch of de Sitter space in the presence of a timelike boundary. We impose that the conformal class of the induced metric and the trace of the extrinsic curvature, $K$, are fixed at the boundary. We present the thermodynamic structure of de Sitter space subject to these boundary conditions, for static and spherically symmetric configurations to leading order in the semiclassical approximation. In three spacetime dimensions, and taking $K$ constant on a toroidal Euclidean boundary, we find that the spacetime is thermally stable for all $K$. In four spacetime dimensions, the thermal stability depends on the value of $K$. It is established that for sufficiently large $K$, the de Sitter static patch subject to conformal boundary conditions is thermally stable. This contrasts the Dirichlet problem for which the region encompassing the cosmological horizon has negative specific heat. We present an analysis of the linearised Einstein equations subject to conformal boundary conditions. In the worldline limit of the timelike boundary, the underlying modes are linked to the quasinormal modes of the static patch. In the limit where the timelike boundary approaches the cosmological event horizon, the linearised modes are interpreted in terms of the shear and sound modes of a fluid dynamical system. Additionally, we find modes with a frequency of positive imaginary part. Measured in a local inertial reference frame, and taking the stretched cosmological horizon limit, these modes grow at most polynomially.

\end{abstract}

\newpage

\setcounter{tocdepth}{2}
\tableofcontents

\newpage

\section{Introduction}
\label{section: Introduction}

In the absence of an asymptotic spatial or null boundary, the construction of gauge invariant observables subject to the constraints of diffeomorphism redundancies in a theory of gravity becomes a challenging task. One is often led to relational notions \cite{geheniau1956invariants,Komar:1958ymq,Bergmann:1961wa} ({for a review on recent work see \cite{Tambornino:2011vg}}) whereby a given physical phenomenon is measured in relation to some other semiclassical feature. For instance, in inflationary models of the early Universe, we can measure the time-dependence of physical phenomena with respect to the slow classical roll of the background inflaton field. From a more quasilocal perspective, one might imagine decorating spacetime with a worldline \cite{fermi,Pirani:1956tn,Manasse:1963zz,Unruh:1976db}, perhaps slightly thickened into a worldtube, and use this as a reference frame for ambient phenomena. This perspective appears to be of particular value for an asymptotically de Sitter spacetime \cite{Anninos:2011af,Anninos:2017hhn,Anninos:2018svg,Coleman:2021nor,Witten:2023qsv,Blacker:2023oan,Loganayagam:2023pfb,Kudler-Flam:2023qfl}, where not only are the Cauchy spatial slices potentially compact, but quasilocal entities are moreover surrounded by a cosmological event horizon rendering most of the expanding portion of spacetime physically obscure. A drawback of the relational approach is that it often necessitates the presence of a semiclassical feature in spacetime, making the general picture away from the semiclassical or perturbative regime difficult to control. 

An alternative, complementary, route may be to study the gravitational theory on a manifold endowed with a quasi-auxiliary timelike boundary $\Gamma$, much like we do when considering gravitational physics in anti-de Sitter space, and try to make sense of general relativity in such a setting. This setup has been the focus of recent work in mathematical relativity \cite{Friedrich:1998xt,Anderson_2008,Sarbach:2012pr,Fournodavlos:2020wde,An:2021fcq,Fournodavlos:2021eye}, accompanied by \cite{Witten:2018lgb,Anninos:2022ujl,Anninos:2023epi} (see also \cite{Figueras:2011va, Adam:2011dn} for related work). Of particular interest to our work is the proposal  of \cite{Anderson_2008,An:2021fcq} that  in four spacetime dimensions certain conformal data along $\Gamma$ lead to a well-posed initial boundary value problem. In \cite{Anderson_2008,An:2021fcq} it is further established that generically, the Dirichlet problem in general relativity---whereby one fixes the induced metric along $\Gamma$---suffers from potential existence and non-uniqueness issues for both Euclidean and Lorentzian signature. Concretely, the conformal boundary conditions of interest fix the conformal class of the induced metric, $[g_{mn}|_\Gamma]_{\text{(conf)}}$, and the trace of the extrinsic curvature, $K$, along $\Gamma$ whilst also specifying standard Cauchy data along a spalike surface $\Sigma$ intersecting $\Gamma$ at its boundary.

In this paper, we explore the conformal boundary conditions of \cite{Anderson_2008,An:2021fcq} for general relativity with a positive cosmological constant $\Lambda$ \cite{Spradlin:2001pw,Anninos:2012qw,Galante:2023uyf}. We consider the problem in both Euclidean and Lorentzian signatures. In Euclidean signature, our main goal is to compute the semiclassical approximation of the gravitational path integral and consider its interpretation from the point of view of Euclidean gravitational thermodynamics \cite{Gibbons:1976ue,Gibbons:1977mu}. In the absence of a boundary, the natural Euclidean geometry for general relativity with $\Lambda>0$ is the sphere \cite{Gibbons:1976ue}, void of any external data such as the size of a thermal circle, and one is led to path integrate fields on top of it. In adding a boundary to our Euclidean manifold, as pointed out in \cite{Hayward:1990zm,Coleman:2021nor,Banihashemi:2022jys} among other places, one has the possibility of providing a novel perspective to the sphere path integral and its rich though elusive physical content \cite{Anninos:2020hfj}.\footnote{More speculatively, as suggested in \cite{Anninos:2021eit}, the presence of boundaries in the $\Lambda>0$ Euclidean path integral might be necessitated in parallel to the non-perturbative necessity of boundaries on the two-dimensional worldsheet of closed strings \cite{Polchinski:1994fq} upon trying to make sense of the sum over topologies. From a Lorentzian perspective, we may imagine the creation of a long-lived heavy particle from the vacuum. These events are Boltzmann suppressed but can occur. In the semiclassical limit, such timelike features become parametrically long-lived and may warrant a treatment involving timeilke boundaries.} In Lorentzian signature, one is led to the question of dynamical features of de Sitter space, known to be dynamically stable at the classical level \cite{Friedrich:1986qfi}, in the presence of a timelike boundary. There are two timelike surfaces of particular interest. One of these is the worldline limit, whereby the spatial size of $\Gamma$ becomes small in units of $\Lambda$. The other is the cosmological horizon limit, whereby $\Gamma$ approaches the cosmological de Sitter horizon. The former limit is of interest in describing the theory of a quasilocal entity, whilst the latter is of interest if one wishes to describe physics from the perspective of a stretched horizon \cite{Susskind:1993if,banks2012holographic,Shaghoulian:2021cef,Shaghoulian:2022fop,Narovlansky:2023lfz}.

\subsubsection*{Organisation and summary of main results}

In section \ref{sec: general}, we present the general framework, and provide an explicit definition of the conformal boundary conditions. As noted, our gravitational theory is endowed with a positive cosmological constant $\Lambda = +(D-1)(D-2)/2\ell^2$ in $D$ spacetime dimensions.

In sections \ref{sec: 3d conformal thermo} and  \ref{sec: 4d}, we consider the problem in Euclidean signature for $D=3$ and $D=4$ spacetime dimensions respectively. The boundary of our manifold is taken to have an $S^1\times S^{D-2}$ topology. In the standard treatment of semiclassical black hole thermodynamics \cite{Gibbons:1976ue} with Dirichlet boundary conditions, one defines the canonical ensemble by fixing the size of the boundary $S^1$ to be the inverse temperature $\beta$ and the radius of the spatial sphere to be fixed to some size   $\frakr$. In our treatment, we will instead fix the conformal class of the boundary metric. As such, we fix a conformal version of the inverse temperature, $\tilde{\beta} \equiv \beta/\frakr$. The other boundary data we fix is the trace of the extrinsic curvature, $K$. We refer to this ensemble as the conformal canonical ensemble. Gravitational solutions in the conformal canonical ensemble include patches with no horizons, referred to as {pole patches}, patches with cosmological horizons, referred to as {cosmic patches}, and patches with black hole horizons, referred to as {black hole patches}. Upon tuning $\tilde{\beta}$ and $K$, one finds that the space is filled with a pure de Sitter spacetime, and we refer to this as a {pure de Sitter patch}.

The complete thermal phase space of static and spherically symmetric solutions at a given  $\tilde{\beta} > 0$ and $K \ell \in \mathbb{R}$ is provided in both $D=3$ and $D=4$ spacetime dimensions. Below we summarise the main results, with an emphasis on the conformal thermodynamics of the pure de Sitter patch.

\textbf{Three-spacetime dimensions.} Our analysis in $D=3$ spacetime dimensions naturally builds on recent developments on dS$_3$ \cite{Coleman:2021nor,Coleman:2020jte,Shyam:2021ciy,Anninos:2021ihe}. We find that upon imposing conformal boundary conditions:

\begin{itemize}
\item Both a pole and a cosmic patch exist at any value of $\tilde{\beta}$ and $K\ell$.

\item The entropy of the cosmic patch is given by the Gibbons-Hawking entropy of the cosmological horizon, and the specific heat is positive for all $\tilde{\beta}$ and $K\ell$.

\item The thermodynamic quantities take the form of a two-dimensional conformal field theory. Viewed as such, we identify a $c$-function
\begin{equation}
   \frak{c}_{\text{conf}} = \frac{3\ell}{4G_N}\left(\sqrt{K^2\ell^2+4}-K\ell\right)\,,
\end{equation}
which decreases monotonically as one goes from the worldline limit, where $K\ell\to-\infty$, to the stretched horizon, where $K\ell \to +\infty$.

\item There is a phase transition at $\tilde{\beta}_c = 2\pi$ (for all values of $K \ell$). At $\tilde{\beta}>\tilde{\beta}_c$, the cosmic patch is the thermally preferred solution. In the stretched horizon limit, the cosmic patch is thermally stable, while in the worldline limit it is metastable.

\end{itemize}

The thermodynamic picture in $D=3$ contrasts that of the canonical ensemble, obtained by imposing Dirichlet boundary conditions. In the canonical ensemble, the specific heat of the cosmic patch is always negative.

\textbf{Four-spacetime dimensions.} We now summarise the situation in $D=4$ spacetime dimensions, which naturally builds on previous work in \cite{Hayward:1990zm,Wang:2001gt,Draper:2022ofa, Banihashemi:2022jys,Banihashemi:2022htw}. We find that upon imposing conformal boundary conditions:
\begin{itemize}
\item A pole patch solution exists for any value of $\tilde{\beta}$ and $K\ell$. On the other hand, cosmic/black hole horizon patch solutions only exist for certain values of $\tilde{\beta}$ and $K\ell$. When they do exist, we identify three distinct solutions for a given $\tilde{\beta}$ and $K \ell$---one pole patch and two horizon patches,  which can be of the cosmic or black hole type.

\item The entropy of the solutions with horizons is always given by their respective horizon area formulas. The solution with the larger horizon area always has positive specific heat, while the one with a smaller horizon area always has negative specific heat.

\item When the cosmic patches have positive specific heat, one can take a large temperature limit. In this limit, the entropy goes as $ N_{\text{d.o.f.}}/\tilde{\beta}^{2}$, and thus resembles that of a conformal field theory in three dimensions. We identify
\begin{equation}
    N_{\text{d.o.f.}} = \frac{32 \pi^3 \ell^2}{81 G_N} \left(\sqrt{K^2\ell^2+9}-K\ell\right)^2 \,.
\end{equation}
Further taking the large $K\ell$ limit of the above expression yields $N_{\text{d.o.f.}} \approx \tfrac{8 \pi^3}{G_N K^2}$, a behaviour identified in \cite{Anninos:2023epi} for black holes in Minkowski space subject to conformal boundary conditions.

\item We identify pure dS$_4$ patches with positive specific heat. These solutions exist when the tube is positioned sufficiently near the cosmological horizon, starting from $\frakr_{\text{tube}} \approx 0.259 \ell$. Depending on $K\ell$, these solutions are either metastable or globally stable. The pure dS$_4$ patch is thermally stable in the stretched horizon limit, at least among the spherically symmetric sector. Near the worldline regime, pure dS$_4$ patches  have negative specific heat. The full phase diagram is presented in figure \ref{fig: phase4d}. 
\end{itemize}
We can contrast the thermodynamic behaviour above to that of the canonical thermal ensemble stemming from Dirichlet boundary conditions \cite{Hayward:1990zm,Wang:2001gt,Draper:2022ofa, Banihashemi:2022jys,Banihashemi:2022htw}. In the latter, the specific heat of the pure dS$_4$ patch is always negative.

In section \ref{sec: dynamics}, we consider the four-dimensional Lorentzian picture. The theory is placed on a manifold with timelike boundary of $\mathbb{R} \times S^{2}$ topology. In the Lorentzian case, one must further supply standard Cauchy data along the initial time spatial slice $\Sigma$. Employing the Kodama-Ishibashi method \cite{Kodama:2000fa}, we present the linearised gravitational dynamics about the pure de Sitter solution. The linearised solutions split into vector and scalar modes concerning their transformation properties under $SO(3)$. We are mainly interested in two limits: one in which the boundary is close to the worldline observer, which we call the worldline limit; and a second one, in which the boundary becomes close to the cosmological horizon, which we call the stretched horizon limit. In each case, the main results of our analysis are:
\begin{itemize}
\item In the worldline limit of the cosmic patch, we retrieve a set of modes that approximate the quasinormal modes of the static patch \cite{Lopez-Ortega:2006aal}, whilst also uncovering a family of modes in the scalar sector with a negative imaginary part. The latter modes have a Minkowskian analogue uncovered in \cite{Anninos:2023epi}.
\item In the stretched horizon limit, our modes degenerate into a variety of modes. The low-lying vector modes match a set of modes identified in \cite{Anninos:2011zn} as a type of linearised shear mode for an incompressible non-relativistic Navier-Stokes equation. The scalar modes take either the form of a sound mode with diverging speed of sound as we approach the horizon limit, or a pair of modes with $\omega \ell = \pm i$. We provide an understanding of these two modes from a purely Rindler perspective and note that in a local inertial frame the exponential behavior becomes polynomial. 
\end{itemize}

Additional technical details are provided in the various appendices.

\section{General framework} \label{sec: general}

We consider vacuum solutions to general relativity with positive cosmological constant $\Lambda= +(D-1)(D-2)/2\ell^2$ in $D=3$, and $D=4$ spacetime dimensions. In Euclidean signature, the action $I_E$ is given by
\begin{equation}
	I_E \= -\frac{1}{16 \pi G_N} \int_{\mathcal{M}} d^D x \sqrt{\text{det} \, g_{\mu\nu}} \,\left(R-2\Lambda\right) - \frac{\alpha_{\text{b.c.}}}{(D-1) 8 \pi G_N} \int_\Gamma d^{D-1}x \sqrt{\text{det}\,g_{mn}} \, K  \, , \label{euclidean_action}
\end{equation}
where $G_N$ is the Newton's constant, $\Gamma = \partial \mathcal{M}$, $g_{mn}$ denotes the induced metric at $\Gamma$, and the trace of the extrinsic curvature $K$ is given by
\begin{equation}\label{eqn: def of K}
	K \= g^{mn} K_{mn} \, , \qquad K_{mn} \= \frac{1}{2}\mathcal{L}_{\hat{n}}g_{mn} \, .
\end{equation}
Here, $\hat{n} \= \hat{n}^\mu \partial_\mu$ is an outward pointing unit normal vector associated with the boundary, and $\mathcal{L}_{\hat{n}}$ denotes a Lie derivative with respect to $\hat{n}^\mu$. We adopt the notation in which Greek indices $\mu  \= 0,...,D-1$ are used for spacetime indices and $m \=0,...,D-2$ are used for spacetime indices tangent to the boundary.

The constant $\alpha_{\text{b.c.}}$ in (\ref{euclidean_action}) depends on the choice of boundary conditions. In most of the paper we will consider conformal boundary conditions, in which we fix the conformal class of the induced metric and the trace of the extrinsic curvature at the boundary, 
\begin{equation}
	\text{Conformal boundary conditions} \quad : \quad \{\left[g_{mn}|_\Gamma\right]_\text{conf} \, , K|_\Gamma \} \quad=\quad \text{fixed} \, .
\end{equation}
With this set of boundary conditions, the initial boundary value problem in general relativity is proven to be well-posed in Euclidean signature \cite{Anderson_2008,Witten:2018lgb} and conjectured to be well-posed in Lorentzian signature \cite{An:2021fcq, Anninos:2022ujl}. This is in contrast to Dirichlet or Neumann boundary conditions, where general relativity does not permit a well-posed initial boundary value problem for generic boundary data. On occasion, it will be useful to contrast results between different boundary conditions. One has
\begin{equation}
\alpha_{\text{b.c.}} = \begin{cases}
1 & \, \, \text{for conformal boundary conditions} \,, \\
(D-1) & \, \, \text{for Dirichlet boundary conditions} \,, \\
\frac{(4-D) (D-1)}{2} & \, \, \text{for Neumann boundary conditions} \,,
\end{cases}
\end{equation}
to ensure the variational principle is well-defined. Note that for Dirichlet boundary conditions, this gives the standard Gibbons-Hawking-York term \cite{Gibbons:1976ue, yorkbdy} and that conformal and Neumann boundary conditions have the same action in $D=3$ \cite{Odak:2021axr}.

Regardless of the choice of the boundary term, the equations of motion satisfied in the interior manifold are the Einstein field equations
\begin{equation}\label{eqn: Einstein field}
	R_{\mu \nu} - \frac{1}{2}g_{\mu\nu}R + \Lambda  g_{\mu \nu}  \= 0 \, .
\end{equation}

\subsection{Conformal thermodynamics}

Following \cite{Anninos:2023epi}, we would like to study the thermodynamic behaviour of solutions subject to conformal boundary conditions, but now in the presence of $\Lambda>0$.

For this, we take the topology of the boundary to be $S^1 \times S^{D-2}$, and consider the following boundary data,
\begin{equation}\label{eqn: euclidean conf bdry cond}
	\left.ds^2\right|_{\Gamma} \= e^{2 \boldsymbol{\omega}}\left(d\tau^2 + \frakr^2 d\Omega^2_{D-2}\right)\, , \qquad\qquad K \= \text{constant} \, ,
\end{equation}
where $\boldsymbol{\omega}$ is an unspecified function that in principle could depend on boundary coordinates,\footnote{In most of the paper, we will consider solutions that have constant $\boldsymbol{\omega}=\omega$, but it is also possible to find non-static solutions where $\boldsymbol{\omega}$ depends on the boundary coordinates. For instance, consider the case where the metric in the bulk is purely de Sitter in $D=3$, with radius $\ell$. Fixing the trace of the extrinsic curvature at $r=\frakr$ to be a constant, $K$, imposes the following differential equation on $\boldsymbol{\omega}$ (assuming it is only a function of boundary time $\tau$),
\begin{equation}\label{key}
    \frakr^2\partial_\tau^2\boldsymbol{\omega}  = 1-(\frakr\partial_\tau \boldsymbol{\omega})^2-\frac{2\frakr^2e^{2\boldsymbol{\omega}}}{\ell^2}-K\frakr e^{\boldsymbol{\omega}} \sqrt{1-(\frakr\partial_\tau \boldsymbol{\omega})^2 - \frac{\frakr^2 e^{2\boldsymbol{\omega}}}{\ell^2}} \,.
\end{equation}
This equation may have solutions apart from $\boldsymbol{\omega} = \omega$ constant (a preliminary numerical analysis indeed suggests solutions periodic in $\tau$). A similar phenomenon occurs in Lorentzian signature, for higher dimensional cases, and also for $\Lambda = 0$. We leave a full analysis of these solutions for future work. Just as a simple concrete example, one can consider Euclidean de Sitter solutions in $D=3$. One solution is simply given by choosing $\boldsymbol{\omega}$ constant, which gives $K \ell$ as in (\ref{poleK}). One could also consider the (Euclidean) de Sitter slicing. In this case, the bulk metric can be conveniently written as 
\begin{equation}
    \frac{ds^2}{\ell^2} = d\rho^2 + \frac{\sin^2\rho}{\frakr^2\cosh^2\tau/\frakr} \left( d\tau^2 + \frakr^2 d\phi^2 \right) \,,
\end{equation}
which at a constant $\rho=\rho_0$, has the same boundary conditions as in (\ref{eqn: euclidean conf bdry cond}), but now $\boldsymbol{\omega}$ depends on $\tau$. The trace of the extrinsic curvature at the boundary is given by $K \ell = 2 \cot \rho_0$, so one can choose $\rho_0$ so that both solutions have the same boundary data. Nonetheless, the time-symmetric spatial slice with $\tau = 0$, which has vanishing extrinsic curvature, has a different proper area than the constant $\boldsymbol{\omega}$ solution. Moreover, the above metric is not periodic in $\tau$. A Lorentzian version of these configurations is obtained by taking $\tau \to i t$. A subset of solutions to (\ref{key}) will appear at the linearised level, and we analyse them in appendix \ref{sec: l0/l1 modes}. These additional solutions need not spoil the uniqueness properties of the Lorentzian conformal boundary conditions, as they have distinguishable Cauchy data.} 
and $d\Omega^2_{D-2}$ is the round metric of the unit $(D-2)$-sphere. The Euclidean time coordinate $\tau \sim \tau + \beta$ parameterises the $S^1$ factor. The parameter $\frakr$ characterises the size of the $S^{D-2}$. Given that only the conformal class of the metric is specified, only the dimensionless parameter $\tilde{\beta}\equiv\beta/\frakr$ is geometrically meaningful. 

To define the conformal canonical ensemble, we consider a partition function $\mathcal{Z}(\tilde{\beta},K)$ as
\begin{equation}\label{eqn: defining partition function}
	\mathcal{Z}(\tilde{\beta}, K) \,\equiv\, \sum_{g_{\mu\nu}^*} e^{-I_E [g_{\mu\nu}^*]} \, ,
\end{equation}
where $g_{\mu\nu}^*$ are Euclidean metrics satisfying the Einstein field equation \eqref{eqn: Einstein field} and obeying the boundary conditions \eqref{eqn: euclidean conf bdry cond}. Note that if there is more than one solution with the same boundary data, we sum all of them. 

According to the Gibbons-Hawking prescription \cite{Gibbons:1976ue}, we interpret $\mathcal{Z}(\tilde{\beta},K)$ as a leading contribution to the thermodynamics partition function in the $G_N \rightarrow 0$ limit. Since we do not fix the Euclidean time periodicity but rather the dimensionless ratio $\tilde{\beta}$, we interpret this as a thermal system in a conformal canonical ensemble at a fixed conformal temperature $\tilde{\beta}^{-1}$.

Given the partition function $\mathcal{Z}(\tilde{\beta},K)$, one can compute different thermodynamic quantities. For instance, the conformal energy, conformal entropy, and specific heat at fixed $K$ are given by
\begin{equation}\label{eqn: defining thermo quant}
	E_\text{conf} \,\equiv\, - \left.\partial_{\tilde{\beta}}\right|_K \log{\mathcal{Z}} \, , \qquad \mathcal{S}_\text{conf} \,\equiv\, \left.\left(1-\tilde{\beta}\partial_{\tilde{\beta}}\right)\right|_K \log{\mathcal{Z}} \, , \qquad C_K \,\equiv\, \left.\tilde{\beta}^2 \partial^2_{\tilde{\beta}}\right|_K \log{\mathcal{Z}} \, .
\end{equation}

\noindent \textbf{Regular Euclidean solutions.} For certain ranges of $\tilde{\beta}$ and $K$, the Einstein field equation may give rise to a solution $g_{\mu\nu}^*$ which contains a Euclidean horizon. In analogy to the Dirichlet case, requiring the solution to be regular at the horizon fixes its size in terms of $\tilde{\beta}$ and $K$. We will consider $g^*_{\mu\nu}$ that are both static and spherically symmetric, taking the explicit form
\begin{equation}\label{mini}
	ds^2 \= e^{2\omega} \left(\frac{f(r)}{f(\frakr)}d\tau^2 + \frac{dr^2}{f(r)} + r^2 d\Omega^2_{D-2}\right)\, ,
\end{equation}
where $\omega$ is a constant and $f(r)$, a function of $r$ only. (Although it would be interesting to explore the existence of saddles subject to conformal boundary conditions with less restrictive symmetry properties than (\ref{mini}), we will postpone such an analysis to future work.) Note that at the boundary $r=\frakr$ with an outward\footnote{It is also possible to consider solutions with the inward normal vector. In such cases, the trace of the extrinsic curvature at the boundary will have an additional minus sign. For our cases of interest, such choice will give rise to the pole and black hole patches in the next two sections.} normal vector $\hat{n} \= \sqrt{f(r)}\partial_r$, 
\begin{equation}\label{eqn: conf bdry cond example}
	\left.ds^2\right|_{\Gamma} \= e^{2 \omega}\left(d\tau^2 + \frakr^2 d\Omega^2_{D-2}\right)\, , \qquad\qquad \left.K\right|_{\Gamma}  \= \frac{f'(\frakr)}{2 e^{\omega}\sqrt{f(\frakr)}} + \frac{D-2}{e^{\omega}\frakr}\sqrt{f(\frakr)} \, ,
\end{equation}
so this metric satisfies conformal boundary conditions \eqref{eqn: euclidean conf bdry cond}. We further assume that $f(r)$ has a simple root at $r=r_+$, so that
\begin{equation}
	f(r) \= (r-r_+)f'(r_+) + \mathcal{O}\left(r-r_+\right)^2\, .
\end{equation}
Then, close to $r_+$, its near horizon geometry (to leading order) is given by,
\begin{equation}
	ds^2 \= \rho^2 \left(\frac{f'(r_+)}{2\sqrt{f(\frakr)}}\right)^2 d\tau^2 + d\rho^2 + e^{2\omega}r_+^2 d\Omega^2_{D-2} \, ,
\end{equation} 
where $\rho \,\equiv\, 2 e^\omega \sqrt{\frac{r-r_+}{f'(r_+)}}$. This geometry has a conical singularity near $r\=r_+$ unless one identifies $\tau \,\sim\, \tau + \beta$ with
\begin{equation}\label{eqn: ref beta}
	\beta \= \frac{4\pi\sqrt{f(\frakr)}}{\left|f'(r_+)\right|} \, .
\end{equation} 
Thus, regularity near the horizon fixes the horizon radius $r_+$ in terms of boundary data $\tilde{\beta}$ and $K$.

\section{dS$_3$ conformal thermodynamics}\label{sec: 3d conformal thermo}

We first study conformal thermodynamics of three-dimensional gravity with $\Lambda = + 1/\ell^2 > 0$. A family of static Euclidean solutions to \eqref{eqn: Einstein field} is given by
\begin{equation}\label{eqn: euclidean sol 3D}
	ds^2 \= e^{2\omega} \left(\frac{f(r)}{f(\frakr)}d\tau^2 + \frac{dr^2}{f(r)} + r^2 d\phi^2 \right)\, , \qquad\qquad f(r) \= \frac{\frakr_{\text{c}}^2 - e^{2\omega} r^2}{\ell^2} \, ,
\end{equation} 
where $\tau \,\sim\, \tau \, + \, \beta$ and $\phi \,\sim\, \phi \, + \, 2\pi$. The choice for this particular parameterisation of the solution will soon become evident. The parameter $\omega$ is an unspecified constant and directly controls the physical size of the boundary, namely $\frakr_\text{tube} \= e^{\omega}\frakr$. The cosmological horizon is located at $r\=e^{-\omega}\frakr_{\text{c}}$ and has a physical radius $\frakr_{\text{c}}>0$. There is no black hole horizon in the present setup.\footnote{Note, however, that related theories of three-dimensional gravity with a gravitational Chern-Simons term admit black hole solutions \cite{Anninos:2009jt}. Moreover, quantum black holes have recently been constructed in dS$_3$ \cite{Emparan:2022ijy,Panella:2023lsi}.} However, for $\frakr_{\text{c}} \,\neq\, \ell$, there is a conical defect located at the origin $r\=0$.

Note that one can recover the standard dS static patch coordinates $\left(\tau_\text{static}, r_\text{static}, \phi_\text{static}\right)$ 
via the identification
\begin{equation}
	\tau_\text{static} \= \frac{e^\omega \tau}{\sqrt{f(\frakr)}} \, , \qquad\qquad r_\text{static} \= e^\omega r \, , \qquad\qquad  \phi_\text{static} \= \phi \, .
\end{equation}

It is straightforward to verify that due to the choice of parameterisation \eqref{eqn: euclidean sol 3D}, the metric automatically satisfies the first boundary condition in \eqref{eqn: euclidean conf bdry cond} at $r\=\frakr$. Requiring that the trace of the extrinsic curvature at the boundary is constant, further fixes the parameter  $\omega$ in terms of the boundary data,
\begin{equation} \label{omega}
	e^{2\omega} \= \frac{\frakr_{\text{c}}^2}{2\frakr^2}\frac{\sqrt{K^2\ell^2+4}\pm K\ell}{\sqrt{K^2\ell^2+4}} \,,
\end{equation}
where the $\pm$ corresponds to two different spacetime regions of interest, which we discuss below. We also note that, assuming that $\omega$ does not depend on $\tau$, we find that \eqref{eqn: euclidean sol 3D} is the most general solution to the Einstein field equation in three dimensions.

\begin{figure}[h!]
        \centering
        \includegraphics[scale=0.25]{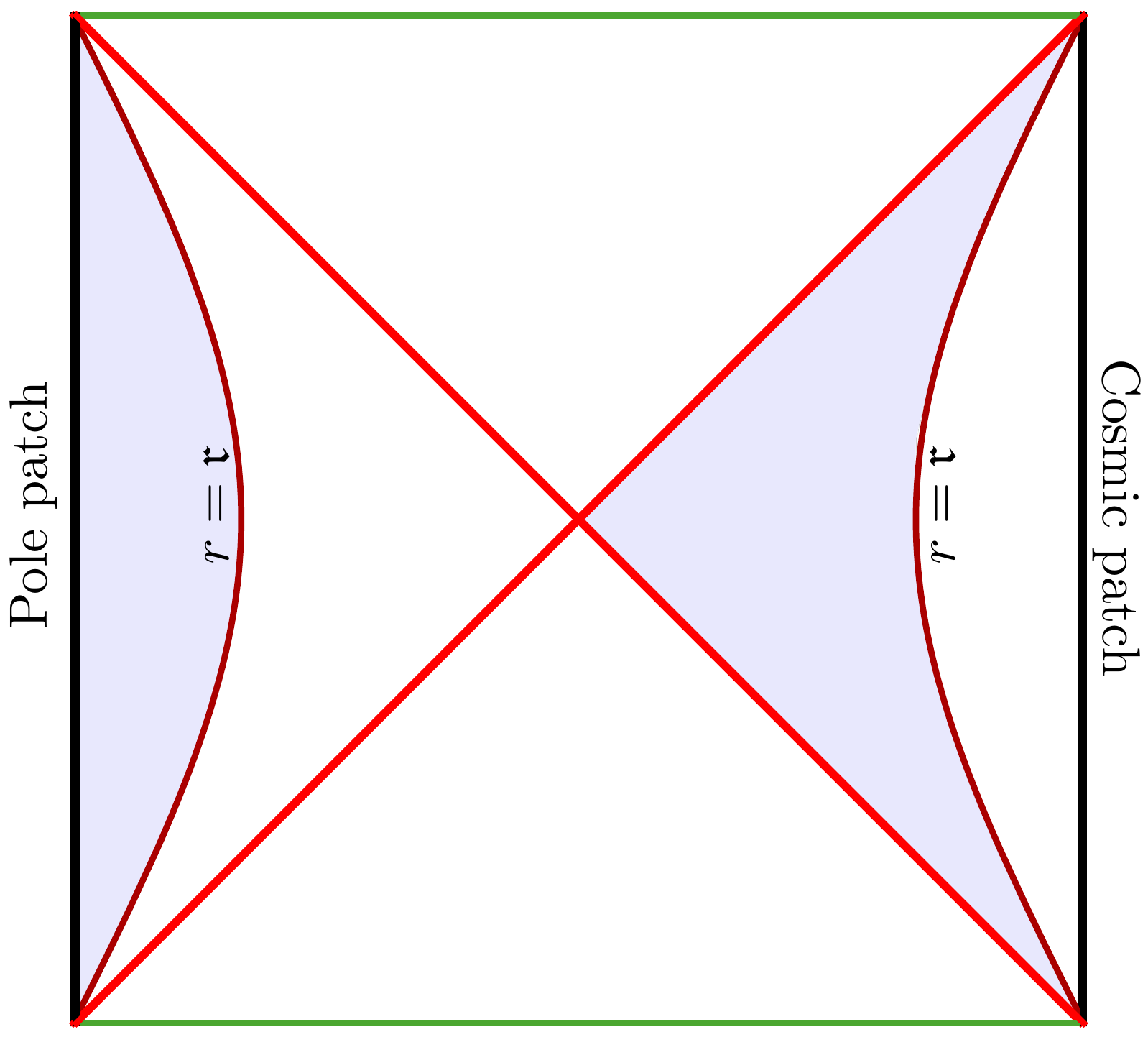}      
                \caption{Penrose diagram of dS space, with a timelike boundary at $r=\frakr$. On the left static patch, the shaded region corresponds to the pole patch, while on the right, it corresponds to the cosmic patch.} \label{fig: cosmic3d}
\end{figure}

We consider two classes of solutions, which we call the pole and the cosmic patch \cite{Coleman:2021nor}. The first one is a patch of spacetime which does not contain the cosmological horizon, while the second one does. In Lorentzian signature they would correspond to the regions shown in figure \ref{fig: cosmic3d}. Below we study the two solutions and their corresponding thermodynamic quantities, separately.

\subsection{Pole patch} 
In the first class of solutions, the cosmological horizon is fixed to be at $\frakr_{\text{c}} \= \ell$, leading to the absence of the conical defect. The spacetime region of interest is $r \,\in\, \left[0,\frakr\right]$, with the boundary at $r \= \frakr \,\leq\, \ell$. We call this spacetime the pole patch of dS. This solution can be obtained by choosing the minus sign in (\ref{omega}).

Imposing that the boundary has a constant trace of the extrinsic curvature $K$ leads to
\begin{equation} \label{poleK}
	K\ell \= \frac{\ell^2-2\frakr_\text{tube}^2}{\frakr_\text{tube}\sqrt{\ell^2-\frakr^2_\text{tube}}} \,,
\end{equation}
which  can be inverted  to obtain
\begin{equation}
	\frakr_\text{tube}^2 \= \frac{\ell^2}{2}\frac{\sqrt{K^2\ell^2+4}-K\ell}{\sqrt{K^2\ell^2+4}} \, .
\end{equation}
The dimensionless parameter $K\ell \,\in\, \mathbb{R}$ controls the size of the boundary. For $K\ell \,\to\, +\infty$ the boundary locates near the origin, whilst for $K\ell \,\to\, -\infty$ it is located near the cosmological horizon. When $K \ell \=0$, the boundary is located exactly at $\frakr_\text{tube} \= \ell/\sqrt{2}$.

Since the cosmological horizon is not part of the pole patch, the parameter $\tilde{\beta}$ is free, and there is a pole patch solution for all values of $\tilde{\beta}$ and $K$.

\textbf{Pole patch thermodynamics.} By evaluating \eqref{euclidean_action} with $\alpha_{\text{b.c.}}=1$ and $D=3$ on the pole patch solution, the on-shell Euclidean action in terms of the boundary data becomes,
\begin{equation}\label{eqn: thermal dS action 3d}
	I_E^{(\text{pole})} \= -\frac{\tilde{\beta}\ell}{16G_N}\left(\sqrt{K^2\ell^2+4}-K\ell\right) \, .
\end{equation}
Since the action depends linearly on $\tilde{\beta}$, one immediately finds that $\mathcal{S}_\text{conf}\=C_K\=0$ and that
\begin{equation}\label{eqn: thermal dS energy 3d}
	E_\text{conf} \= \frac{I_E^{(\text{pole})}}{\tilde{\beta}} =  -\frac{\ell}{16G_N}\left(\sqrt{K^2\ell^2+4}-K\ell\right) \, ,
\end{equation}
which is independent of $\tilde{\beta}$. Note that small fluctuations of the energy can be written as,
\begin{equation}
	\delta E_\text{conf} \= \frac{\frakr_\text{tube}^2}{8G_N} \,\delta K \,.
\end{equation}
Following \cite{York:1986it}, we treat the pole patch of dS as a reference configuration and the on-shell action \eqref{eqn: thermal dS action 3d} as a subtraction term. Therefore, the conformal energy \eqref{eqn: thermal dS energy 3d} plays the role of a vacuum energy. From now onwards, we will compute subtracted quantities such that the energy of the pole patch solution with trace of extrinsic curvature $K$ vanishes.

\subsection{Cosmic patch}
We now consider the second class of static geometries \eqref{eqn: euclidean sol 3D}, which contain the cosmological horizon and hence are dubbed as cosmic patches of dS. 
This is achieved by choosing the plus sign on \eqref{omega} and considering the region $r \,\in\, \left[\frakr,e^{-\omega}\frakr_{\text{c}}\right]$. As a consequence, the conical defect at $r=0$ is not part of the cosmic patch. 

Regularity of the geometry near the cosmological horizon imposes that the inverse conformal temperature of the cosmic patch is given by
\begin{equation}\label{eqn: cosmo ds3 beta}
	\tilde{\beta} \= \frac{2\pi\ell\sqrt{\frakr_{\text{c}}^2 - \frakr_{\text{tube}}^2}}{\frakr_{\text{c}} \frakr_\text{tube}} \,, 
\end{equation}
which is always greater than zero. The conformal temperature $\tilde{\beta}^{-1}$ becomes zero as the boundary approaches the origin. On the other hand, the conformal temperature diverges to infinity as the boundary approaches the cosmological horizon.

Requiring that the boundary has a constant trace of the extrinsic curvature $K$, fixes
\begin{equation}\label{eqn: cosmo ds3 K}
	K \ell \= -\frac{\frakr_{\text{c}}^2-2\frakr_\text{tube}^2}{\frakr_\text{tube}\sqrt{\frakr_{\text{c}}^2-\frakr^2_\text{tube}}} \,,
\end{equation}
which can take any real value. Contrary to the pole patch, the limit of $K \ell$ going to positive and negative infinity now corresponds to the limit of the boundary approaching the cosmological horizon and the origin, respectively. This is expected as the normal vector now points in the opposite direction. 

Using \eqref{eqn: cosmo ds3 beta} and \eqref{eqn: cosmo ds3 K}, we can express $\frakr_{\text{c}}$ and $\frakr_\text{tube}$ in terms of the boundary data $\tilde{\beta}$ and $K\ell$,
\begin{equation}
	\frakr_{\text{c}} \= \frac{\pi\ell}{\tilde{\beta}}\left(\sqrt{K^2\ell^2+4}-K\ell\right) \, , \qquad\qquad \frakr_\text{tube} \= \frac{\sqrt{2}\pi\ell}{\tilde{\beta}}\sqrt{\frac{\sqrt{K^2\ell^2+4}-K\ell}{\sqrt{K^2\ell^2+4}}} \, .
\end{equation}
Interestingly, both $\frakr_{\text{c}}$ and $\frakr_\text{tube}$ depend linearly on the conformal temperature $\tilde{\beta}^{-1}$. This fact implies that the cosmological horizon $\frakr_{\text{c}}$ is a monotonically increasing function of the conformal temperature, 
which contrasts with the Dirichlet problem, where one finds an opposite behaviour, see appendix \ref{sec:3d_Dirichlet}. 

\begin{figure}[h!]
        \centering
        \includegraphics[width=9 cm]{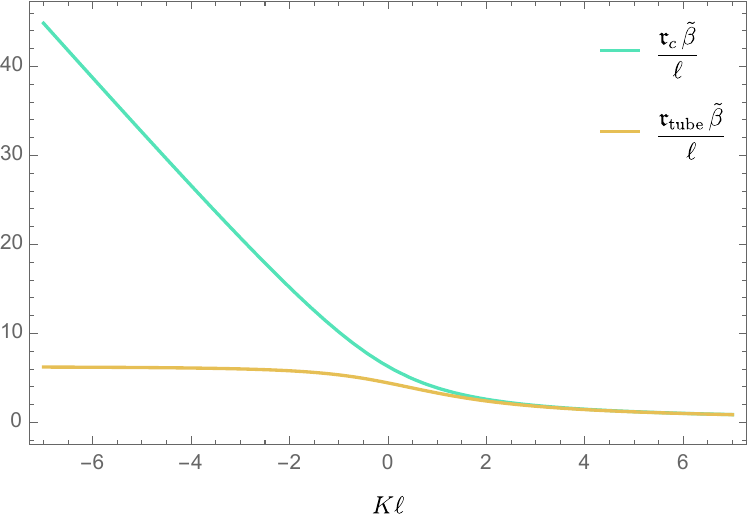}      
                \caption{Plot of $\frac{\frakr_{\text{c}}\tilde{\beta}}{\ell}$ (green) and $\frac{\frakr_\text{tube}\tilde{\beta}}{\ell}$ (yellow) as functions of $K\ell$. For $K \ell \to \infty$, the tube becomes very close to the cosmological horizon; for $K \ell \to - \infty$, the tube approaches $\frac{\frakr_\text{tube}\tilde{\beta}}{\ell} = 2\pi$.} \label{fig: rc rt versus K 3d}
\end{figure}

Additionally, for any positive $\tilde{\beta}$ and real $K$, one always finds that $0\,<\,\frakr_\text{tube}\,<\, \frakr_{\text{c}}$. We show $\frac{\frakr_{\text{c}} \tilde{\beta}}{\ell}$ and $\frac{\frakr_\text{tube} \tilde{\beta}}{\ell}$ as functions of $K\ell$ in figure \ref{fig: rc rt versus K 3d}. 

\begin{figure}[h!]
        \centering
        \includegraphics[width=9 cm]{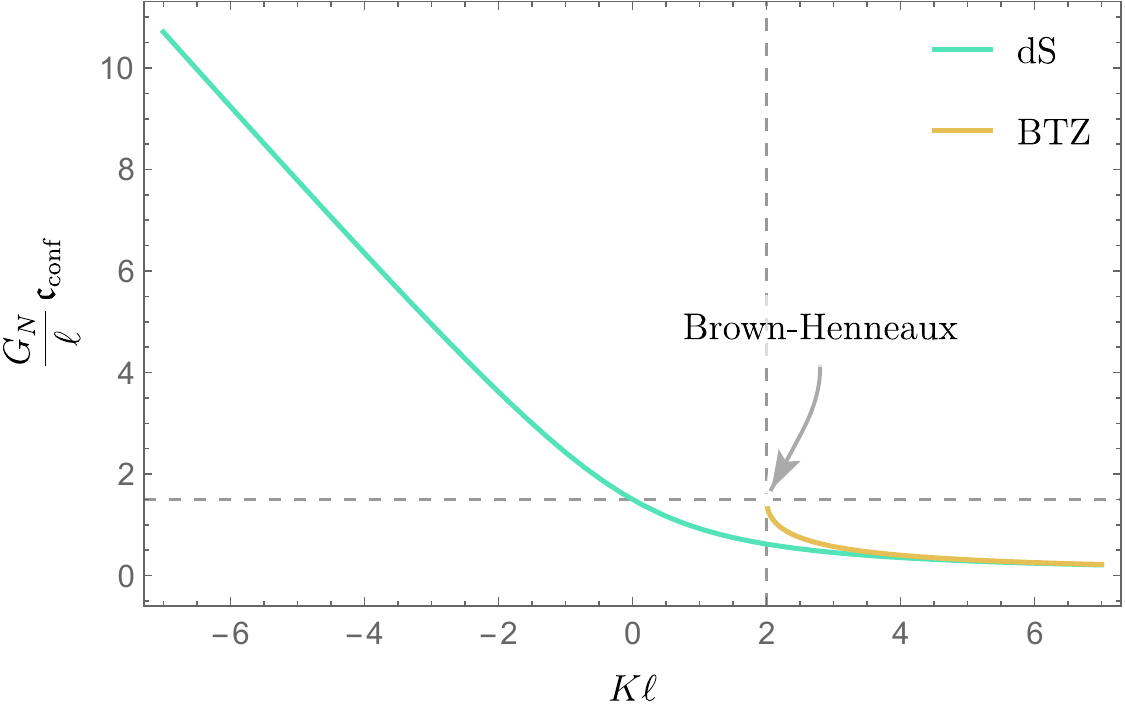}      
                \caption{The quantity $\frak{c}_{\text{conf}}$ as a function of $K \ell$. This central charge decreases monotonically as a function of $K \ell$. As comparison, we included the plot of the analogous $\frak{c}_{\text{conf}}^{\text{BTZ}}$, for the BTZ black hole with conformal boundary conditions. The dS and AdS radii are chose to be equal. In the limit where $K \ell \to \infty$ both central charges coincide, as the tube in both cases gets very close to the horizon and they both exhibit Rindler behaviour. In dashed lines, we show the position of the conformal boundary in the AdS case, where we recover the Brown-Henneaux central charge.
              } \label{fig: central charge 3d}
\end{figure}

 \textbf{Cosmic patch thermodynamics.} To compute thermodynamic quantities, we evaluate the cosmic patch solution on-shell in the Euclidean action \eqref{euclidean_action}. It is useful to define a regulated quantity by subtracting the pole patch action \eqref{eqn: thermal dS action 3d} with the same boundary data. Then,  the associated regulated action corresponding to the cosmic patch solution is given by
\begin{equation}\label{eqn: I_E^cosmo 3d}
	I_{E\text{, reg}}^{(\text{cosmic})} (K) \,\equiv\, I_E^{(\text{cosmic})} - I_E^{(\text{pole})} \= - \frac{\pi \ell}{8G_N}\left(\frac{2\pi}{\tilde{\beta}}-\frac{\tilde{\beta}}{2\pi}\right) \left(\sqrt{K^2\ell^2+4}-K\ell\right)\, .
\end{equation}
We emphasise that the pole action that we subtract has the same trace of extrinsic curvature $K$. We further note that $I_{E\text{, reg}}^{(\text{cosmic})} (K)$ is invariant under $\frac{\tilde{\beta}}{2\pi} \rightarrow - \frac{2\pi}{\tilde{\beta}}$.
The conformal energy and conformal entropy are given by
\begin{equation}\label{eqn: conformal energy entropy cosmic 3d}
	\begin{cases}
		E_\text{conf} &=\, \frac{\pi \ell}{8 G_N \tilde{\beta}} \left(\frac{2\pi}{\tilde{\beta}} + \frac{\tilde{\beta}}{2\pi}\right)\left(\sqrt{K^2\ell^2+4}-K\ell\right) \, , \\
		\mathcal{S}_\text{conf} &=\, \frac{\pi^2 \ell}{2G_N \tilde{\beta}} \left(\sqrt{K^2\ell^2+4}-K\ell\right) \, .
	\end{cases}
\end{equation}
The entropy $\mathcal{S}_\text{conf}$ agrees with the Gibbons-Hawking entropy $A_\text{horizon}/4G_N$ of the cosmological horizon. 
We can also compute the specific heat at constant $K$, which is given by
\begin{equation}\label{eqn: specific heat cosmic 3d}
	C_K \= \frac{\pi^2\ell}{2G_N\tilde{\beta}} \left(\sqrt{K^2\ell^2+4}-K\ell\right) \, .
\end{equation}
The specific hear is positive for all allowed values of $\tilde{\beta}$ and $K$, which means that this configuration is thermally stable under small thermal fluctuations. This is in contrast to the specific heat of the cosmic patch with Dirichlet boundary conditions, which is always negative, as reviewed in appendix \ref{sec:3d_Dirichlet}.
Moreover, note that $C_K$ grows linearly with the conformal temperature. This behaviour resembles that of a two-dimensional conformal field theory at finite temperature. This observation can be sharpened 
upon expressing $\tilde{\beta}$ in terms of $E_\text{conf}$, whereby the conformal entropy and specific heat can be written as
\begin{equation}\label{eqn: entropy cardy}
	\mathcal{S}_\text{conf} \= 2 \pi \sqrt{\frac{\frak{c}_{\text{conf}}}{3}\left(E_\text{conf}-\frac{\frak{c}_{\text{conf}}}{12}\right)} \, \ , \qquad\qquad \, C_K  \= \frac{2\pi^2\frak{c}_{\text{conf}}}{3\tilde{\beta}}  \, ,
\end{equation}
where
\begin{equation}\label{eqn: central charge 3d}
	\frak{c}_{\text{conf}} \equiv \frac{3\ell}{4G_N}\left(\sqrt{K^2\ell^2+4}-K\ell\right) \, ,
\end{equation}
which is a monotonically-decreasing function of $K \ell$ displayed in figure \ref{fig: central charge 3d}.\footnote{A similar computation can be done for the BTZ black hole in AdS$_3$ with conformal boundary conditions \cite{Shyam:2021ciy,shaghoulian}. The resulting value for this function is now given by $\frak{c}_{\text{conf}}^{\text{BTZ}} = \frac{3 \ell_{\text{AdS}}}{4 G_N} (K\ell_{\text{AdS}}-\sqrt{K^2\ell_{\text{AdS}}^2-4})$. This is also a monotonically decreasing function of $K \ell_{\text{AdS}}$. It approaches its maximal real value as $K \ell_{\text{AdS}} \to 2$, which corresponds to the AdS conformal boundary, and for which we recover the standard Brown-Henneaux central charge \cite{Brown:1986nw}.} Note that the entropy in \eqref{eqn: entropy cardy} resembles the Cardy formula, describing the growth of states of energy $E_\text{conf}$ in a two-dimensional conformal field theory of central charge $\frak{c}_{\text{conf}}$ \cite{Cardy:1986ie}. By considering large positive/negative $K\ell$ limits, 
\begin{equation}
	\frak{c}_\text{conf} \,\rightarrow\, 
	\begin{cases}
		\frac{3}{2G_N K} \, , & \text{as} \quad K \ell \rightarrow \infty \, , \\
		\frac{3|K|\ell^2}{2G_N} \, , & \text{as} \quad K\ell \rightarrow - \infty \, .
	\end{cases}
\end{equation}

Finally, using \eqref{eqn: entropy cardy}, we find a first-law type of relation in which
\begin{equation} 
	\delta \left(E_\text{conf}-\frac{\frak{c}_{\text{conf}}}{12}\right) \= \tilde{\beta}^{-1}\, \delta \mathcal{S}_\text{conf} - \mu_K \,\delta K\, , 
\end{equation}
where $\mu_K$ can be understood as the chemical potential associated to $K$, 
\begin{equation}
	\mu_K \,\equiv\,  -\frac{\frakr_\text{tube}^2}{8G_N} \, .
\end{equation}

\subsection{Pure dS$_3$ patch}

We now discuss a particular solution, denoted as the pure dS$_3$ solution. The solution arises from tuning the conformal temperature of the system such that $\frakr_{\text{c}} \= \ell$, leading to a pure dS$_3$ with a boundary. For the cosmic patch, this happens at the conformal temperature $\tilde{\beta} = \tilde{\beta}_{\text{dS}}$, which is a function of $K \ell$ given by
\begin{equation}\label{eqn: dS temp 3d}
	\tilde{\beta}_{\text{dS}} (K \ell)  \,\equiv\, \pi \left(\sqrt{K^2\ell^2+4}-K \ell\right) \= 2\pi\left(\frac{2 G_N \, \frak{c}_{\text{conf}}}{3 \ell}\right) \, .
\end{equation}

By using \eqref{eqn: cosmo ds3 K} with $\frakr_{\text{c}} \= \ell$, we can re-express this in terms of the physical size of the boundary $\frakr_\text{tube}$ as
\begin{equation}
	\tilde{\beta}_{\text{dS}}(\frakr_\text{tube}) \= 2\pi \sqrt{\frac{\ell^2}{\frakr_\text{tube}^2}-1} \, .
\end{equation}
We can recover the standard dS temperature when the tube becomes small, 
\begin{equation}
	\tilde{\beta}_{\text{dS}} \frakr_\text{tube} \,\rightarrow\, 2\pi\ell \qquad\qquad \text{as} \qquad \frakr_\text{tube}\,\rightarrow\, 0\,.
\end{equation}
This is the worldline limit of the solution. It also corresponds to a small conformal temperature. On the other hand, we can consider the stretched horizon limit, which is defined by taking $\frakr_\text{tube} \rightarrow \ell$. This is equivalent to a high conformal temperature limit,
\begin{equation}
	\tilde{\beta}_{\text{dS}} \,\rightarrow\, 0 \qquad\qquad \text{as} \qquad \frakr_\text{tube}\,\rightarrow\, \ell\,.
\end{equation}

Now we can use \eqref{eqn: conformal energy entropy cosmic 3d} and \eqref{eqn: specific heat cosmic 3d} to calculate the thermodynamic properties of pure dS$_3$. The conformal entropy and specific heat at constant $K$ are equal and independent of $\frakr_\text{tube}$,
\begin{equation}
	\mathcal{S}_\text{conf} \= C_K \= \frac{\pi \ell}{2G_N} \, .
\end{equation}
We stress again that the specific heat $C_K$ is positive for all values of $\frakr_{\text{tube}}$.
The conformal energy reads
\begin{equation}
	E_\text{conf}  \= \frac{\frak{c}_{\text{conf}}}{12} \left(1-\frac{\frakr_\text{tube}^2}{\ell^2}\right)^{-1} \, .
\end{equation}
Note that in the worldline limit, the energy reduces to the energy $\frak{c}_{\text{conf}}/12$.
One may wonder why the energy does not go to zero as we take worldline limit. The reason is that we are considering subtracted energies. 

To reproduce the full static patch thermodynamics from a cosmic patch solution, we must include the pole patch to complete a full static patch. Note that this pole patch is not the same that we are using to define the regulated action, as this one has the same $\tilde{\beta}$ but the opposite trace of the extrinsic curvature, $-K$. Now it is straightforward to check that, for $\tilde{\beta}\=\tilde{\beta}_{\text{dS}}$, a combination of a pure de Sitter patch with $K$ and a pole patch with $-K$ (without the subtraction terms) indeed reproduces the Gibbons-Hawking result,
\begin{equation}
 \mathcal{Z}^{\text{(cosmic)}} (\tilde{\beta}_{\text{dS}},K) \mathcal{Z}^{\text{(pole)}} (\tilde{\beta}_{\text{dS}},-K) = \exp \frac{\pi \ell}{2 G_N}\, ,
\end{equation}
as the right-hand side is the exponential of the Gibbons-Hawking entropy for the de Sitter horizon. For the energy, we can define a regulated action for this pole patch with opposite $K$,
\begin{equation}
	I_{E\text{, reg}}^{(\text{pole})}(-K) \, \equiv \, I_{E}^{(\text{pole})}(-K) - I_{E}^{(\text{pole})}(K) \= - \frac{\tilde{\beta} K\ell^2}{8G_N} \, .
\end{equation}

In the pure dS$_3$ solution, the conformal energy of the pole patch with $-K$ is negative and it is exactly opposite to the conformal energy of the cosmic patch with $K$ such that 
\begin{equation}
	\left.\left(E_\text{conf}^{(\text{pole})} (-K) - \frac{\frak{c}_\text{conf}}{12}\right) + \left(E_\text{conf}^{(\text{cosmic})} (K) - \frac{\frak{c}_\text{conf}}{12}\right)\right|_{\tilde{\beta} \= \tilde{\beta}_{dS}} \= 0,
\end{equation}
as we expect for the full static patch of de Sitter space.

\subsection{Phase diagram}

\begin{figure}[h!]
        \centering
        \includegraphics[width=9 cm]{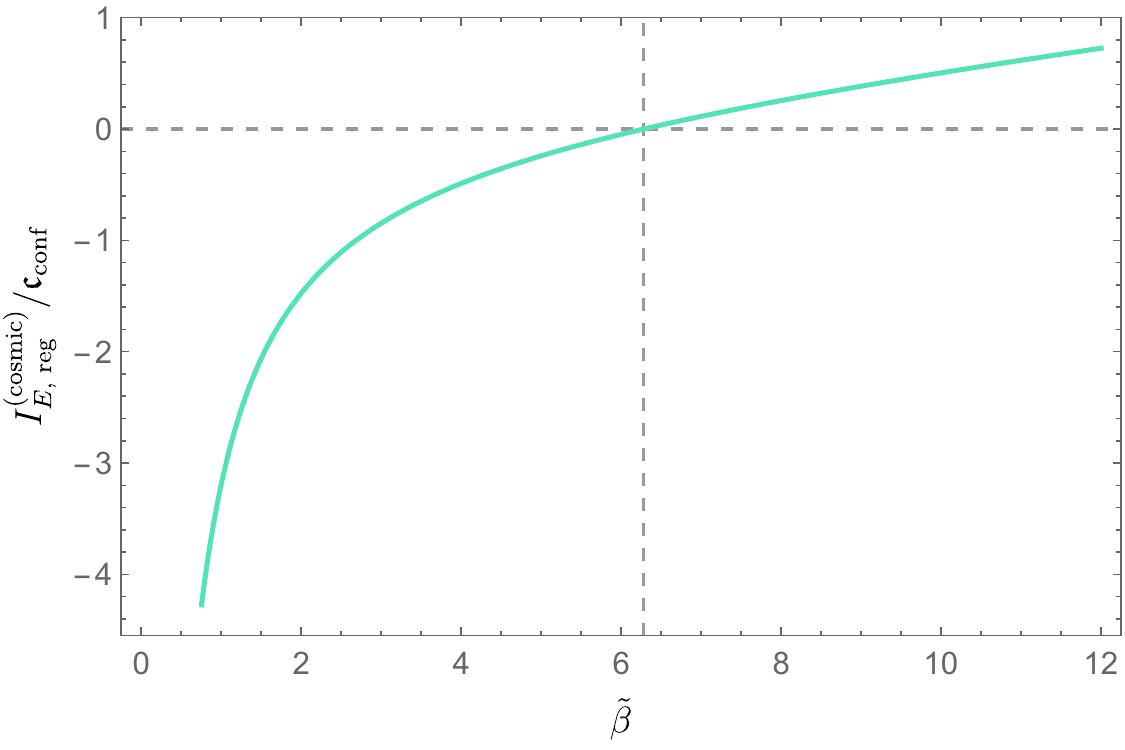}      
                \caption{The regulated on-shell action for the cosmic patch solution, as a function of boundary data $\tilde{\beta}$. The dashed vertical line indicates the critical inverse temperature $\tilde{\beta}_c=2\pi$.} \label{fig: free energy 3d}
\end{figure}

Now we can combine the results from pole patch and cosmic patch thermodynamics. Recall that in $D=3$, both solutions exist for all values of $\tilde{\beta}$ and $K$. This means that one may write the partition function of the total system in the semiclassical limit as
\begin{equation}
	\mathcal{Z} (\tilde{\beta},K) \= \text{max}\left(e^{-I_{E\text{, reg}}^{(\text{cosmic})}(\tilde{\beta},K)}, 1\right) e^{-I_E^{(\text{pole})}(\tilde{\beta},K)} \, ,
\end{equation}
where $I_E^{(\text{pole})}(\tilde{\beta},K)$ and $I_{E\text{, reg}}^{(\text{cosmic})}(\tilde{\beta},K)$ are given by \eqref{eqn: thermal dS action 3d} and \eqref{eqn: I_E^cosmo 3d}, respectively. The sign of $I_{E\text{, reg}}^{(\text{cosmic})}(\tilde{\beta},K)$ therefore determines which configuration is stable/meta-stable. It can be shown that, independently of $K\ell$, there is a critical inverse temperature $\tilde{\beta}\= 2\pi \equiv \tilde{\beta}_c $, for which $I_E^{(\text{cosmic})}(\tilde{\beta},K)$ changes sign, see figure \ref{fig: free energy 3d}. There is a first-order phase transition at $\tilde{\beta} = \tilde{\beta}_c$.

\begin{itemize}
    \item In the low-temperature regime with $\tilde{\beta} \,>\, \tilde{\beta}_c$, $I_{E\text{, reg}}^{(\text{cosmic})}$ is positive for all values of $K \ell$ which implies that the pole patch is thermodynamically favoured. Given it has positive specific heat, the cosmic patch is then metastable.
    \item In the high-temperature regime with $\tilde{\beta} \,<\, \tilde{\beta}_c $, $I_{E\text{, reg}}^{(\text{cosmic})}$ becomes negative, so it becomes thermodynamically favored and a stable configuration.
\end{itemize}

It is interesting to analyse the phase structure of conformal thermodynamics at fixed $K \ell$. Consider the conformal energy along the path of lowest free energy at fixed $K \ell$. Using \eqref{eqn: thermal dS energy 3d} and \eqref{eqn: conformal energy entropy cosmic 3d} and expressing them in terms of the central charge $\mathfrak{c}_{\text{conf}}$, we find that
\begin{equation}
	E_\text{conf} \= 
	\begin{cases}
		\frac{\mathfrak{c}_{\text{conf}}}{12}\left(1+\frac{\tilde{\beta}_c^2}{\tilde{\beta}^2}\right) \, ,& \qquad \tilde{\beta}\,<\, \tilde{\beta}_c\, , \\
		0\, , & \qquad \tilde{\beta} \,\geq\,  \tilde{\beta}_c \, .
	\end{cases}
\end{equation}
The discontinuity of $E_\text{conf}$ at $\tilde{\beta} \= \tilde{\beta}_c$ reflects a first-order phase transition. 

For the conformal entropy, we find a behaviour similar to the Hawking-Page transition of the AdS black hole \cite{Hawking:1982dh, Witten:1998zw}. For temperatures lower than $\tilde{\beta}^{-1}_c$, the conformal entropy is zero. There is a discontinuity in the entropy at the critical temperature $\tilde{\beta}^{-1}_c$, after which, in the high temperature regime, the entropy is precisely given by the Gibbons-Hawking entropy. 

\textbf{Pure dS$_3$ phase structure.} For the pure dS$_3$ solution, we  must  constrain the inverse temperature to the dS inverse temperature \eqref{eqn: dS temp 3d}. In this case, $K \ell\=0$ corresponds to $\frakr_{\text{tube}} \= \ell/\sqrt{2}$.

\begin{figure}[h!]
        \centering
        \includegraphics[scale=0.35]{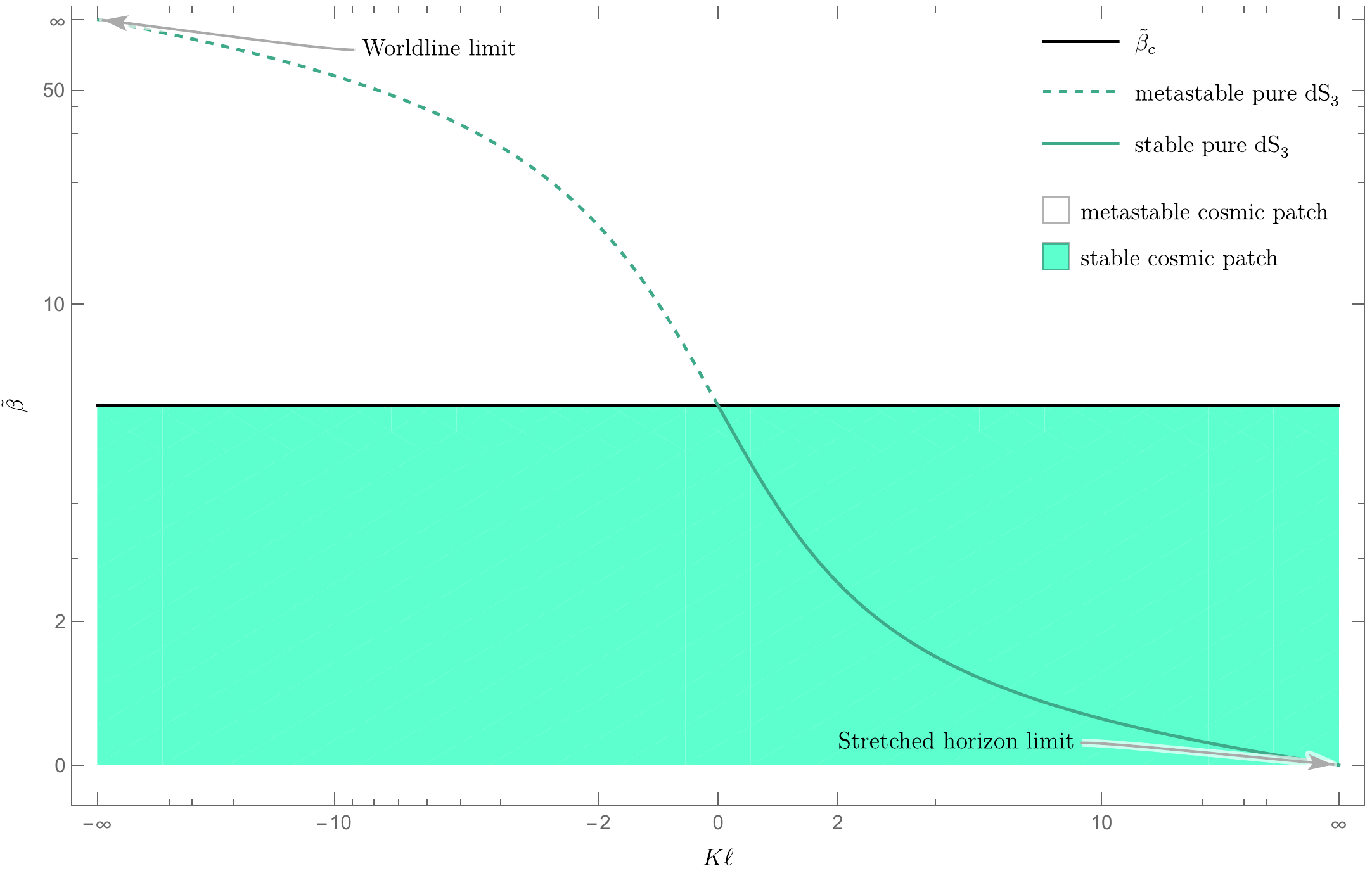}      
                \caption{Phase diagram of conformal dS$_3$ thermodynamics. In each point of the diagram, there co-exist a pole and a cosmic patch solution. There is a critical conformal inverse temperature at $\tilde{\beta}_c=2\pi$, marked in black. For $\tilde{\beta}< \tilde{\beta}_c$ (shaded in green), the cosmic patch is the most favourable configuration. For $\tilde{\beta}> \tilde{\beta}_c$ (in white), the cosmic patch is metastable. The darker green curve shows pure stable (solid) and metastable (dashed) dS$_3$ solutions, that follow relation (\ref{eqn: dS temp 3d}). Worldline and stretched horizon limits of pure dS$_3$ are further indicated. }\label{fig: phase3d}
\end{figure}

\begin{itemize}
    \item $K \ell \,>\,0$ implies that $\frak{c}_{\text{conf}} \,<\,\frac{3\ell}{2G_N}$. In this regime, the dS temperature is higher than the critical temperature, $\tilde{\beta}_{\text{dS}}\,<\,\tilde{\beta}_c$. 
As a consequence, in this regime, the pure dS$_3$ has free energy lower than the pole patch and hence is thermodynamically favoured. 
    \item For $K \ell <0$ or $\frak{c}_{\text{conf}}>\frac{3\ell}{2G_N}$, we find that $\tilde{\beta}_{\text{dS}}\,>\,\tilde{\beta}_c$, 
so the pure dS solution is only metastable. Lastly, at $\frak{c}_{\text{conf}} \= \frac{3\ell}{2G_N}$, the phase transition and dS$_3$ temperature coincide. 
\end{itemize}
The full phase diagram, including the curve of pure dS$_3$ solutions, is depicted in figure \ref{fig: phase3d}.

\subsection{A two-sphere perspective}

As a final remark before moving on to the four-dimensional case, we consider a two-sphere rather than toroidal boundary topology. As our coordinate system, we take
\begin{equation}
    ds^2 \= \frac{\ell^2 d\rho^2}{\ell^2-\rho^2} + \frac{\ell^2-\rho^2}{\ell^2}\left(d\theta^2 + \sin^2\theta d\phi^2\,\right) \, ,
\end{equation}
where for the full three-sphere we have $\rho\,\in\, \left(-\ell,\ell\right)$, $\theta \in (0, \pi)$, and $\phi \,\sim\, \phi+2\pi$. We take the conformal boundary to be located at constant $\rho \= \rho_0$. The induced metric has the conformal structure of the unit $S^2$ metric. Requiring further that the boundary has a constant $K$ fixes
\begin{equation}
    K\ell \= -\frac{2\rho_0}{\sqrt{\ell^2-\rho_0^2}} \, ,
\end{equation}
where we have used a unit normal vector $\hat{n} \= \frac{\ell}{\sqrt{\ell^2-\rho^2}}\partial_\rho$. By computing the on-shell action \eqref{euclidean_action} for  this solution, we can approximate the path integral as
\begin{equation}
    \mathcal{Z}[{S^2}] \approx  \left( \frac{ K \ell - 2i}{\sqrt{K^2 \ell^2 + 4}}  \right)^{\frac{i}{3} \times \frac{3  \ell}{2 G_N}}~. 
\end{equation}
The above expression is real valued. Unlike the thermodynamic expression (\ref{eqn: conformal energy entropy cosmic 3d}) and (\ref{eqn: entropy cardy}), the above expression does not immediately take the form of the two-sphere path integral of a two-dimensional conformal field theory of central charge $\frak{c}_{\text{conf}}$. A similar observation will hold for Euclidean AdS$_3$ with a two-sphere boundary subject to conformal boundary conditions.

\section{dS$_4$ conformal thermodynamics} \label{sec: 4d}

In this section, we study conformal thermodynamics of four-dimensional gravity with $\Lambda \= + 3/\ell^2$ for the following family of static and spherically symmetric Euclidean solutions
\begin{equation}\label{eqn: euclidean sol 4D}
	ds^2 \= e^{2\omega} \left(\frac{f(r)}{f(\frakr)}d\tau^2 + \frac{dr^2}{f(r)} + r^2 \left( d\theta^2 + \sin^2{\theta} d\phi^2 \right) \right)\, , \qquad f(r) \= 1- \frac{2\mu}{r} - e^{2\omega} \frac{r^2}{\ell^2}\, ,
\end{equation} 
where $\tau \,\sim\, \tau + \beta$, $\theta \,\in\, (0, \pi)$, and $\phi \,\sim\, \phi + 2\pi$. Similarly to the three-dimensional case, the parameter $\omega$ controls the size of the boundary, namely $\frakr_{\text{tube}} \= e^\omega \frakr$. 

In $D=4$, one further has the Euclidean Schwarzschild-de Sitter solution corresponding to a black hole placed inside the cosmological horizon. The parameter $\mu$ is related to the size of the cosmological horizon $\frakr_{\text{c}}$ through
\begin{equation}
	\mu \= \frac{e^{-\omega}\frakr_{\text{c}}}{2}\left(1-\frac{\frakr_{\text{c}}^2}{\ell^2}\right) \, .
\end{equation}
For $\mu \= 0$, we obtain an empty de Sitter solution with  cosmological horizon at $\frakr_{\text{c}} \= \ell$. For $\mu \,>\, 0$, there is a black hole horizon located at $r\=e^{-\omega}\frakr_{\text{bh}}$ with horizon radius $\frakr_{\text{bh}}$. The cosmological horizon in this case is smaller than the one without the black hole, $\frakr_{\text{c}}\, <\, \ell$. Note that the size of the two horizons are related by
\begin{equation}\label{eqn: relation between rbh and rc}
	\frakr_{\text{bh}} \= \frac{1}{2}\left(\sqrt{4\ell^2-3\frakr_{\text{c}}^2}-\frakr_{\text{c}}\right)  \, .
\end{equation}
Both horizon sizes coincide when $\frakr_{\text{c}} \= \frakr_{\text{bh}} \= \frac{\ell}{\sqrt{3}}$. This is known as the Nariai radius, which also serves as a lower bound of $\frakr_{\text{c}}$.  For $\mu \,<\,0$, there is a naked singularity located at $r \= 0$, and the cosmological horizon is greater than the de Sitter length, $\frakr_{\text{c}} \,>\,\ell$.

As in dS$_3$, we note that the de Sitter static patch coordinates $\left(\tau_\text{static}, r_\text{static}, \theta_\text{static},\phi_\text{static}\right)$ can be recovered 
via the rescaling
\begin{equation}
	\tau_\text{static} \= \frac{e^\omega \tau}{\sqrt{f(\frakr)}} \, , \qquad\qquad r_\text{static} \= e^\omega r \, , \qquad\qquad  \theta_\text{static} \= \theta\, , \qquad\qquad  \phi_\text{static} \= \phi \, .
\end{equation}

We now study these solutions on a four-manifold with an $S^1 \times S^2$ boundary subject to the conformal boundary data \eqref{eqn: euclidean conf bdry cond}. Solving the Einstein equation in terms of boundary data, we obtain multiple expressions for $e^{2\omega}$. The exact expressions in the different ranges of parameters are provided in appendix \ref{sec: useful formulae for 4d thermo}.

\begin{figure}[h!]
        \centering
        \includegraphics[scale=0.35]{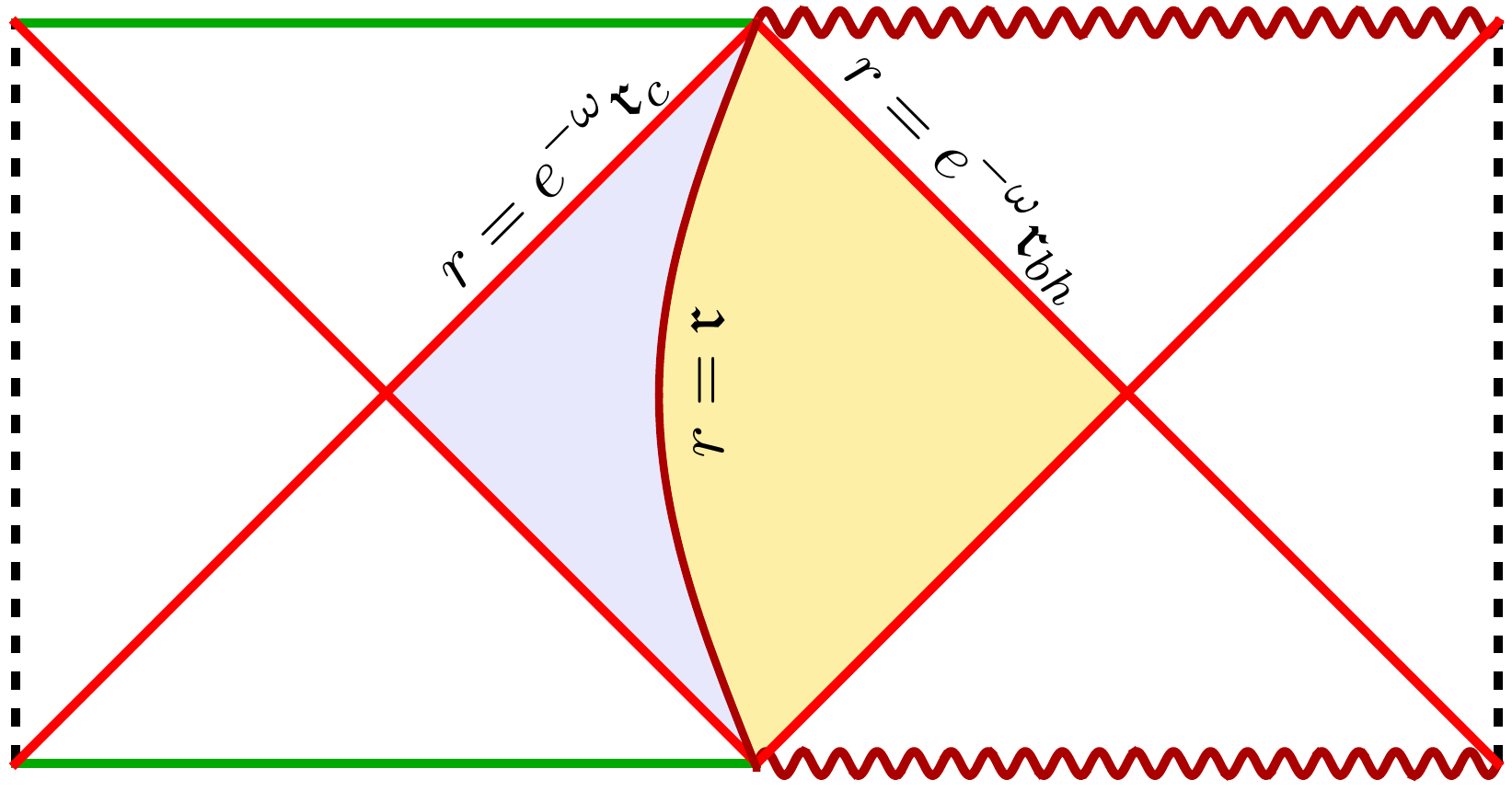}      
                \caption{Penrose diagram of the Schwarzschild de Sitter spacetime. The boundary is given by $r=\frakr$. The shaded blue area corresponds to a cosmic patch, while the yellow one, to a black hole patch. The pole and the pure dS$_4$ patches can be obtained when $\frakr_{\text{c}} = \ell$.} \label{fig: sds}
\end{figure}

There are three different classes of solutions denoted by the pole patch, the black hole patch, and the cosmic patch. Examples of them are displayed in figure \ref{fig: sds}.

\subsection{Pole patch} We begin with the solutions with $\frakr_{\text{c}} \= \ell$ and take the spacetime region of interest to be $r \,\in\, \left[0,\frakr \right]$. As a consequence, the pole patches contain the worldline at $r\=0$ without any horizon.

Imposing that the boundary has a constant trace of the extrinsic curvature $K$ leads to
\begin{equation}
	K\ell \= \frac{2\ell^2- 3\frakr_\text{tube}^2}{\frakr_\text{tube}\sqrt{\ell^2-\frakr_\text{tube}^2}} \, .
\end{equation}
By inverting this equation, we find that the physical size of the boundary can be written as a function of $K \ell$ as
\begin{equation}
	\frakr_\text{tube}^2 \= \ell^2\left(\frac{K^2\ell^2+12-K\ell \sqrt{K^2\ell^2+8}}{2 K^2\ell^2+18}\right) \, .
\end{equation}
The parameter $K\ell$ can take any real value. The limit of large positive and large negative $K\ell$ corresponds to pushing the boundary to the origin and the cosmological horizon, respectively.

Since the pole patches do not contain any horizon, the parameter $\tilde{\beta}$ is free and can take any positive value. Hence, the pole patch exists for all values of $\tilde{\beta}$ and $K$.

\textbf{Pole patch thermodynamics.}
We now evaluate \eqref{euclidean_action} with $\alpha_{\text{b.c.}}=1$ and $D=4$ on the pole patch solution. The on-shell Euclidean action in terms of the boundary data reads,
\begin{equation}\label{eqn: thermal dS action 4d}
	I_E^{(\text{pole})} \= -\frac{\tilde{\beta}\ell^2}{6G_N} \sqrt{\frac{\left(K^2\ell^2+4-K\ell\sqrt{K^2\ell^2+8}\right)\left(K^2\ell^2+12-	K\ell\sqrt{K^2\ell^2+8}\right)}{K^2\ell^2+9}} \, .
\end{equation}
By taking the limit $K\ell \rightarrow \infty$, we find that $I_E^{(\text{pole})} \rightarrow -\frac{4 \tilde{\beta}}{3 G_NK^2}$, retrieving the result in flat space obtained in \cite{Anninos:2023epi}. This is expected as this limit corresponds to a boundary size that is parameterically small compared to the cosmological horizon.

As the action depends linearly on $\tilde{\beta}$, one immediately finds that $\mathcal{S}_\text{conf}\=C_K\=0$ and that
\begin{equation}\label{eqn: thermal dS energy 4d}
	E_\text{conf} \= \frac{I_E^{(\text{pole})}}{\tilde{\beta}} \=  -\frac{\ell^2}{6G_N} \sqrt{\frac{\left(K^2\ell^2+4-K\ell\sqrt{K^2\ell^2+8}\right)\left(K^2\ell^2+12-	K\ell\sqrt{K^2\ell^2+8}\right)}{K^2\ell^2+9}}\, ,
\end{equation}
which is independent of $\tilde{\beta}$. Note that small fluctuations of the energy can be written as
\begin{equation}
	\delta E_\text{conf} \= \frac{\frakr_\text{tube}^3}{3G_N} \,\delta K \,. \label{chemical_pole}
\end{equation}
Curiously, the coefficient in front of $\delta K$ is independent of $\ell$. As in $D=3$, we treat the pole patch of de Sitter as a reference configuration, and the on-shell action \eqref{eqn: thermal dS action 4d} as a subtraction term.

\subsection{Cosmic patch} 

In this section, we consider a class of geometries \eqref{eqn: euclidean sol 4D} which contain the cosmological horizon. We call these cosmic patches of dS. The spacetime region of interest is taken to be $r \,\in\, \left[\frakr,e^{-\omega}\frakr_{\text{c}}\right]$. For $\ell/\sqrt{3} \, < \, \frakr_{\text{c}} \,<\, \ell$, the full solution has a black hole horizon that lies outside the boundary, so it is not present in the cosmic patch. Similarly,  
for $\frakr_{\text{c}} \, > \, \ell$, there is a naked timelike singularity at the origin in Lorentzian signature, which would be associated with the presence of negative energy. Again, since this region is not part of the cosmic patch, we also allow for $\frakr_{\text{c}} \,>\, \ell$.

Regularity of the geometry near the cosmological horizon fixes the conformal temperature of the cosmic patch to be
\begin{equation}\label{eqn: cosmic ds4 beta}
	\tilde{\beta} \= \frac{4\pi \frakr_{\text{c}}\ell\sqrt{\frakr_\text{tube}\ell^2 - \frakr_\text{tube}^3 - \frakr_{\text{c}}\ell^2 + \frakr_{\text{c}}^3}}{\frakr_\text{tube}^{3/2} \left(3 \frakr_{\text{c}}^2-\ell^2\right)} \, ,
\end{equation}
which is always greater than zero. In this case, the conformal temperature $\tilde{\beta}^{-1}$ does not have a lower bound. Specifically, for $\ell/\sqrt{3} \, < \,\frakr_{\text{c}} \, \leq \, \ell$, the zero temperature limit can be reached by setting $\frakr_{\text{c}} \= \ell$ and taking $\frakr_\text{tube}/\ell \rightarrow 0$. For $\frakr_{\text{c}} > \ell$, there are also cosmic patches with zero conformal temperature. They have the boundary located closed to the naked singularity, that is to say $\frakr_\text{tube}/\ell \rightarrow 0$.

The high conformal temperature limit $\tilde{\beta}\rightarrow 0$ can be achieved in different ways, for instance, by taking the near horizon limits, $\frakr_\text{tube} \rightarrow \frakr_{\text{bh}}$ or $\frakr_\text{tube} \rightarrow \frakr_{\text{c}}$.

Setting the trace of the extrinsic curvature at the boundary to be constant leads to
\begin{equation}\label{eqn: cosmo ds4 K}
	K \ell \= -\frac{4\frakr_\text{tube}\ell^2 - 6 \frakr_\text{tube}^3-3\frakr_{\text{c}}\ell^2 + 3 \frakr_{\text{c}}^3}{2\frakr_\text{tube}^{3/2}\sqrt{\frakr_\text{tube}\ell^2-\frakr_\text{tube}^3-\frakr_{\text{c}}\ell^2+\frakr_{\text{c}}^3}} \,.
\end{equation}
There is no upper or lower bound on $K\ell$. The limit of large positive and large negative $K\ell$ correspond to taking the boundary to be near the cosmological horizon and the black hole horizon, respectively. In the case of $\frakr_{\text{c}} \,\geq\, \ell$, there is no black hole and so the large negative limit of $K \ell$ corresponds to taking the boundary to be near the origin.

Unlike the $D=3$ case, we could not find analytic expressions for $\frakr_\text{tube}$ and $\frakr_{\text{c}}$ in terms of the boundary data $\tilde{\beta}$ and $K$, but we relegate some useful analytical expressions to appendix  \ref{sec: useful formulae for 4d thermo}. We later present examples of $\frakr_\text{tube}$ and $\frakr_{\text{c}}$ as functions of $\tilde{\beta}$ at fixed $K\ell$ in figure \ref{fig: radius versus beta}, together with the black hole patch solutions.

\textbf{Cosmic patch thermodynamics.} We start by computing the on-shell action \eqref{euclidean_action} of the cosmic patch. We define a regulated action as the on-shell action of the cosmic patch subtracted by the pole patch action with the same $\tilde{\beta}$ and $K$. By expressing it in terms of $\frakr_{\text{bh}}$ and $\frakr_\text{tube}$, we find that
\begin{equation}\label{eqn: onshell cosmic 4d}
	I_{E\text{, reg}}^{(\text{cosmic})} \= - \frac{\pi \frakr_{\text{c}} \left(4\frakr_\text{tube}\ell^2 - 3 \frakr_{\text{c}}\ell^2 - 3 \frakr_{\text{c}}^3\right)}{3G_N\left(\ell^2-3\frakr_{\text{c}}^2\right)} - I_E^{(\text{pole})} \, ,
\end{equation}
where $I_E^{\text{(pole)}}$ is given by \eqref{eqn: thermal dS action 4d}.

The conformal energy and the conformal entropy of the cosmic patch are
\begin{equation}\label{eqn: conformal energy entropy cosmic 4d}
		E_\text{conf} \= \frac{\frakr_\text{tube}^{3/2}\left(2\frakr_\text{tube}\ell^2-3\frakr_{\text{c}} \ell^2+ 3 \frakr_{\text{c}}^3\right)}{6 G_N\ell \sqrt{\frakr_\text{tube}\ell^2-\frakr_\text{tube}^3-\frakr_{\text{c}}\ell^2+\frakr_{\text{c}}^3}} - \frac{I_E^{(\text{pole})}}{\tilde{\beta}} \, , \qquad\qquad
		\mathcal{S}_\text{conf} \= \frac{\pi \frakr_{\text{c}}^2}{G_N} \, .
\end{equation}
The conformal entropy $\mathcal{S}_\text{conf}$ agrees with the Gibbons-Hawking entropy of the cosmological horizon, $A_\text{horizon}/4G_N$. 
The specific heat at constant $K$ of the cosmic patch is given by
\begin{equation}\label{eqn: spec heat cosmic ds4}
	C_K = \frac{2\pi \frakr_{\text{c}}^2 \left(-\ell^2+3 \frakr_{\text{c}}^2\right)\left(9\frakr_{\text{c}}^2\left(\frakr_{\text{c}}^2-\ell^2\right)^2+16 \frakr_{\text{c}} \left(\frakr_{\text{c}}^2-\ell^2\right)\frakr_\text{tube}\ell^2+8\frakr_\text{tube}^2\ell^4 - 4 \frakr_\text{tube}^4\ell^2\right)}{G_N\left(\ell^2+3\frakr_{\text{c}}^2\right)\left(9 \frakr_{\text{c}}^2\left(\frakr_{\text{c}}^2-\ell^2\right)^2+2\frakr_{\text{c}}\frac{\left(-9\ell^4-10\frakr_{\text{c}}^2\ell^2+15\frakr_{\text{c}}^4\right)}{\left(\ell^2+3\frakr_{\text{c}}^2\right)}\frakr_\text{tube}\ell^2+8\frakr_\text{tube}^2\ell^4-4\frakr_\text{tube}^4\ell^2\right)} \, .
\end{equation}

It is interesting to remark that in the limit where the boundary approaches the cosmological horizon, the specific heat becomes
\begin{equation}\label{eqn: spec heat cosmic 4d}
C_K \,\rightarrow\, \frac{2 \pi \frakr_{\text{c}}^2}{G_N} \,, \qquad \text{as} \quad \frac{\frakr_{\text{tube}}}{\frakr_{\text{c}}} \to 1 \,.
\end{equation}

This positive specific heat is to be contrasted with the negative specific heat that is obtained when Dirichlet boundary conditions are imposed on the cosmic patch \cite{Draper:2022ofa, Banihashemi:2022jys}. We will further discuss this fact when we consider the pure dS$_4$ solutions.

\begin{figure}[h!]
        \centering
        \includegraphics[width=9 cm]{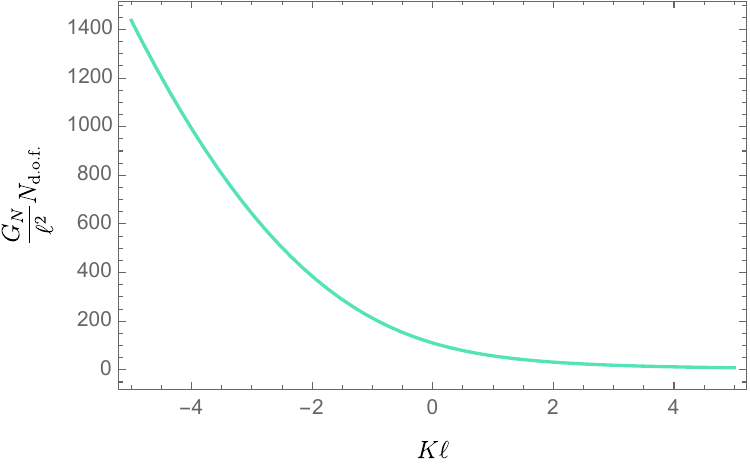}      
                \caption{The number of degrees of freedom $N_{\text{d.o.f.}}$ as a function of $K \ell$. The number of degrees of freedom decreases monotonically as a function of $K \ell$.
              } \label{fig: central charge 4d}
\end{figure}

Interestingly, the high conformal temperature limit of the specific heat \eqref{eqn: spec heat cosmic 4d} at finite $K\ell$ is given by
\begin{equation}
    C_K \,\rightarrow\, \frac{32\pi^3\ell^2}{81\tilde{\beta}^2G_N}\left(\sqrt{K^2\ell^2+9}-K\ell\right)^2\,, \qquad \text{as} \quad\tilde{\beta} \,\rightarrow\, 0 \, .
\end{equation} 
This takes the form of the specific heat of a three-dimensional conformal field theory. Under this interpretation,    the putative number of degrees of freedom goes as
\begin{equation}
    N_\text{d.o.f.} \= \frac{32 \pi^3 \ell^2}{81G_N}\left(\sqrt{K^2\ell^2+9}-K\ell\right)^2 \,,
\end{equation}
which is a monotonically decreasing function of $K\ell$, displayed in figure \ref{fig: central charge 4d}. Further taking $K \ell \to + \infty$ yields $N_\text{d.o.f.} \,\to 8\pi^3 / G_N K^2$ matching the $\Lambda=0$ result in \cite{Anninos:2023epi}.

Finally, we find that the thermodynamic quantities satisfy a first-law type of relation
\begin{equation}
    \delta E_\text{conf} \= \tilde{\beta}^{-1} \delta \,\mathcal{S}_\text{conf} - \mu_K \,\delta K \,,
\end{equation}
where $\mu_K$, similarly to the three-dimensional case, is interpreted as the chemical potential associated to $K$, 
\begin{equation}
    \mu_K \,\equiv\, - \frac{\frakr_\text{tube}^3}{3G_N} - \frac{1}{\tilde{\beta}} \left. \frac{\partial I_E^{(\text{pole})}}{\partial K}\right|_{\tilde{\beta}} \, .
\end{equation}

Note that the first term in $\mu_K$ looks identical to the one that appears for the pole patch in $D=4$, see (\ref{chemical_pole}).

\subsection{Black hole patch} \label{sec: bh patch}
We now consider a class of geometries \eqref{eqn: euclidean sol 4D} which contain the black hole horizon. We refer to these  as black hole patches of dS$_4$. These solutions exist as long as the cosmological horizon radius takes values between the dS length and the Nariai radius, $\frac{\ell}{\sqrt{3}}\,<\,\frakr_{\text{c}}\,<\,\ell$. The spacetime region of interest is then given by $r \,\in\, \left[e^{-\omega}\frakr_{\text{bh}} , \frakr\right]$ where $\frakr_{\text{bh}}$ is related to $\frakr_{\text{c}}$ through \eqref{eqn: relation between rbh and rc}. For convenience, in this section, we will always express $\frakr_{\text{c}}$ in terms of $\frakr_{\text{bh}}$. 

Given that these regions are complementary to the cosmic patches (see figure \ref{fig: sds}) many results for these solutions are closely related to those in the cosmic patch, upon replacing $\frakr_{\text{c}} \to \frakr_{\text{bh}}$. Here we point out the main differences between the two patches and relegate the explicit formulae to appendix \ref{black hole patch app}.

It is easy to obtain the boundary data from the results of the cosmic patch. The inverse conformal temperature is the same as in \eqref{eqn: cosmic ds4 beta}, but with $\frakr_{\text{c}} \to \frakr_{\text{bh}}$ and an extra minus sign. The trace of the extrinsic curvature is minus the expression that appears in \eqref{eqn: cosmo ds4 K}, again with $\frakr_{\text{c}} \to \frakr_{\text{bh}}$. As opposed to the cosmic patch, in this case, the conformal temperature $\tilde{\beta}^{-1}$ has a lower bound $\tilde{\beta}^{-1}_{\text{min}} \= 2\pi$, which occurs in the Nariai limit, by setting $\frakr_\text{tube} \= \ell/\sqrt{3}$ and taking $\frakr_{\text{bh}} \rightarrow \ell/\sqrt{3}$ from below.

Below this conformal temperature, the black hole patch solution does not exist. For larger conformal temperatures, $\tilde{\beta}^{-1} \, > \, \tilde{\beta}^{-1}_{\text{min}}$, there is a one-parameter family of black hole patches.

Regarding the trace of the extrinsic curvature, the limit of $K \ell$ approaching negative infinity corresponds to pushing the boundary to be near the cosmological horizon. For $K\ell$ going to positive infinity, the boundary is pushed near the black hole horizon. The behaviour of $K\ell$ as a function of $\frakr_{\text{c}}$ and $\frakr_\text{tube}$ is exactly opposite to the cosmic patch since the normal vector points in the opposite direction.

\textbf{Black hole patch thermodynamics.} Similarly, thermodynamic quantities can also be obtained from those of the cosmic patch. In particular, the regulated action $I_{E\text{, reg}}^{(\text{bh})}$ is the same as in the cosmic patch, but with $\frakr_{\text{c}} \to \frakr_{\text{bh}}$. The conformal energy is also the same as in \eqref{eqn: conformal energy entropy cosmic 4d}, but with a minus sign in front of the first term and the replacement of $\frakr_{\text{c}} \to \frakr_{\text{bh}}$. The entropy is now given by
\begin{equation}
		\mathcal{S}_\text{conf} \= \frac{\pi \frakr_{\text{bh}}^2}{G_N} \, ,
\end{equation}
which agrees with the Bekenstein-Hawking entropy $A_\text{horizon}/4G_N$ where the horizon in this formula now corresponds to the black hole horizon. The specific heat can also be obtained from \eqref{eqn: spec heat cosmic ds4}, with the replacement $\frakr_{\text{c}} \to \frakr_{\text{bh}}$. In particular, it is also positive as the tube approaches the black hole horizon. Explicit expressions and further interesting limits are shown in appendix \ref{black hole patch app}.

\subsection{Pure dS$_4$ patch}\label{sec: 4d pure dS}

As in the dS$_3$ case,  we can recover the pure dS$_4$ solution for a particular family of conformal temperatures. Using the boundary data of the cosmic patch \eqref{eqn: cosmic ds4 beta} and \eqref{eqn: cosmo ds4 K}, the conditions for having pure dS$_4$ are
\begin{equation}
	\tilde{\beta} \= \tilde{\beta}_{\text{dS}} \,\equiv\, \frac{2 \pi \sqrt{\ell^2-\frakr_\text{tube}^2}}{\frakr_\text{tube}} \, , \qquad\qquad K\ell \= - \frac{2\ell^2-3\frakr_\text{tube}^2}{\frakr_\text{tube}\sqrt{\ell^2-\frakr_\text{tube}^2}} \, .
\end{equation}
From these, it follows that $\frakr_{\text{c}} = \ell$.  One can further solve for $\tilde{\beta}_{\text{dS}}$  in terms of $K \ell$,
\begin{equation}\label{eqn: dS temp 4d}
	\tilde{\beta}_{\text{dS}} \= \frac{\pi}{2} \left(\sqrt{K^2\ell^2+8}-K\ell\right) \, .
\end{equation}

Consider now the worldline limit $\frakr_\text{tube}\,\rightarrow\,0$. The standard dS temperature is recovered when
\begin{equation}
	\tilde{\beta}_{\text{dS}} \frakr_\text{tube} \,\rightarrow\, 2\pi\ell \qquad\qquad \text{as} \qquad \frakr_\text{tube}\,\rightarrow\, 0\,.
\end{equation}
The stretched horizon limit,  corresponds to the high conformal temperature limit in which
\begin{equation}
	\tilde{\beta}_{\text{dS}} \,\rightarrow\, 0 \qquad\qquad \text{as} \qquad \frakr_\text{tube}\,\rightarrow\, \ell\,.
\end{equation}

Now we can use \eqref{eqn: conformal energy entropy cosmic 4d} and \eqref{eqn: spec heat cosmic ds4} to calculate the thermodynamic properties of pure dS$_4$. The conformal entropy and the specific heat at constant $K$ of the pure dS$_4$ are given by
\begin{equation}
	\mathcal{S}_\text{conf} \= \frac{\pi \ell^2}{G_N} \, , \qquad\qquad C_K \= - \frac{2\pi\frakr_\text{tube}\ell^2\left(2\ell^2-\frakr_\text{tube}^2\right)}{G_N\left(\ell^3 - 4 \frakr_\text{tube}\ell^2 + 2 \frakr_\text{tube}^3\right)}\, . \label{ck ds4}
\end{equation}
The conformal entropy is constant regardless of $\frakr_\text{tube}$ and is given by the Gibbons-Hawking entropy of the cosmological horizon, as expected. 

\begin{figure}[h!]
        \centering
        \includegraphics[width=9 cm]{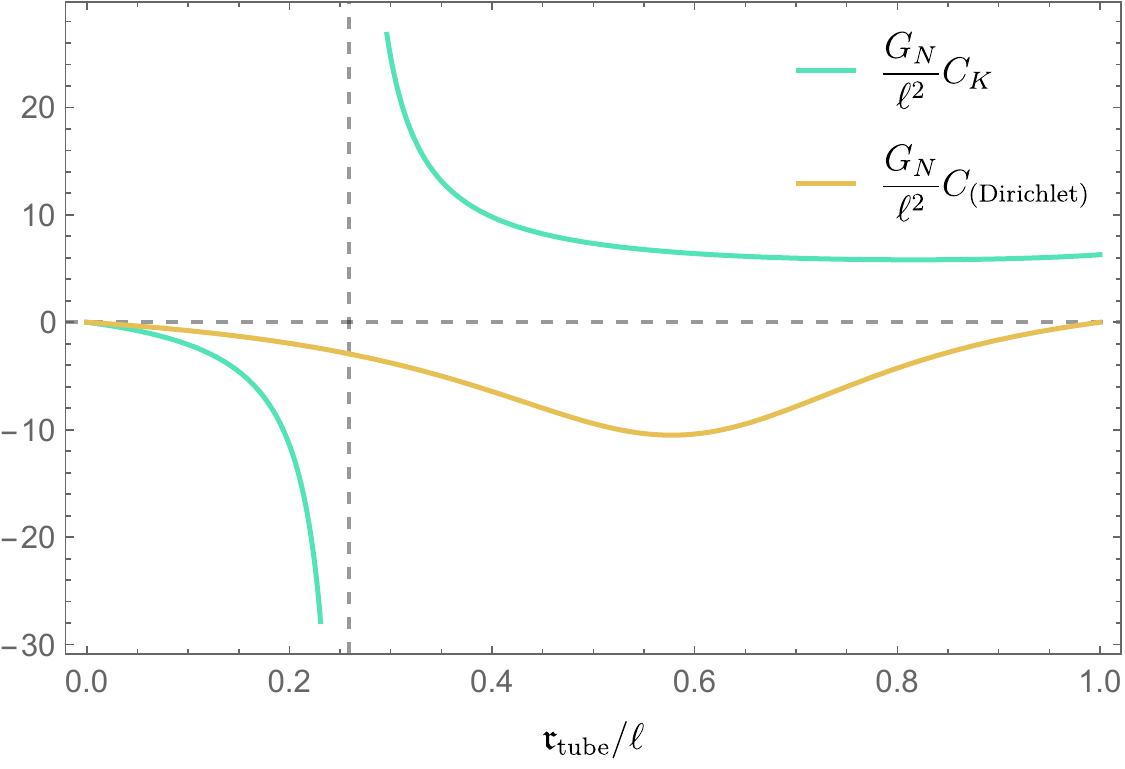}      
                \caption{A plot of the specific heat of the pure de Sitter patch for conformal (green) and Dirichlet (yellow) boundary conditions. For the Dirichlet case, the specific heat is never positive.} \label{fig: specific heat pure dS4}
\end{figure}

It is interesting to note the  behaviour of the specific heat at constant $K$. Close to the worldline, $C_K$ in (\ref{ck ds4}) is negative. In fact, in the worldline limit, the specific heat converges to zero from below, $C_K \,\rightarrow\, - 4\pi \frakr_\text{tube}\ell/G_N$. This is similar to what happens for the Dirichlet case \cite{Draper:2022ofa, Banihashemi:2022jys} where
\begin{equation}
C_{\text{(Dirichlet)}} = - \frac{2\pi \frakr_\text{tube}\ell^2\left(\ell^2-\frakr_\text{tube}^2\right)}{G_N\left(\ell^3 - 2 \frakr_\text{tube}\ell^2 + 2 \frakr_\text{tube}^3\right)}~.
\end{equation}
We plot both specific heats in figure \ref{fig: specific heat pure dS4}. A notable feature is that for the case of conformal boundary conditions, the specific heat diverges as $\frakr_\text{tube} \to \frakr_0$, with $\frakr_0$ given by\footnote{This is the only root to the cubic equation $\ell^3-4\frakr_0\ell^2+2\frakr_0^3 = 0$, satisfying $0<\frakr_0<\ell$, 
\begin{equation}
    \frac{\frakr_0}{\ell} = 2\sqrt{\frac{2}{3}}\cos\left(\frac{1}{3}\cos^{-1}\left(-\frac{3\sqrt{3}}{8\sqrt{2}}\right)-\frac{2\pi}{3}\right) \,.
\end{equation}}
\begin{equation}\label{eqn: K pure dS 4d}
	\frakr_0 \,\approx\, 0.259 \ell \, \qquad \Leftrightarrow \qquad  K \ell \approx -7.202 \,. 
\end{equation}
For $\frakr_\text{tube} \,>\, \frakr_0$, we find that the specific heat is positive and approaches a constant $C_K \,\rightarrow\, 2 \pi \ell^2/G_N$ as we take the stretched horizon limit. 
The pure dS$_4$ patch with a conformal boundary sufficiently close to the de Sitter horizon is thus thermally stable. 


The conformal energy of the pure dS$_4$ is given by
\begin{equation}\label{eqn: conf energy pure ds4}
	E_\text{conf} \= \frac{4\ell^2\left(\ell^2-\frakr_\text{tube}^2\right)}{3\frakr_\text{tube}G_N\sqrt{4\ell^2-3\frakr_\text{tube}^2}} + \frac{\frakr_\text{tube}^2\ell}{3G_N\sqrt{\ell^2-\frakr_\text{tube}^2}}\, ,
\end{equation}
which is positive for all $0\,<\,\frakr_\text{tube}\,<\, \ell$ with a lowest value given by
\begin{equation}\label{eqn: minimum energy pure ds4}
	\left.E_\text{conf}\right|_{\frakr_\text{tube}\=\sqrt{\frac{2}{3}}\ell} \= \frac{4\ell^2}{3\sqrt{3}G_N}.
\end{equation}
In terms of the boundary data, the minimum energy \eqref{eqn: minimum energy pure ds4} is obtained precisely when $K\ell\=0$. Both in the worldline and in the stretched horizon limit, the energy is divergent. In particular, $E_\text{conf} \,\rightarrow\, \frac{2\ell^3}{3\frakr_\text{tube}G_N}$, in the worldline limit and $E_\text{conf} \,\rightarrow\, \frac{\ell^2}{3\sqrt{2}G_N \sqrt{1- \frakr_\text{tube}/\ell}}$, in the stretched horizon limit. 

As in the pure dS$_3$ discussion, one could consider a pole patch which, together with the cosmic patch of pure dS$_4$, completes the full Euclidean static patch geometry, which is the four-sphere. This is achieved by considering a pole patch which has the same $\tilde{\beta}$ but an opposite trace of the extrinsic curvature $-K$. As in $D=3$, it is straightforward to check that at the saddle point level
\begin{equation}
\mathcal{Z}^{\text{(cosmic)}} (\tilde{\beta}_{dS},K) \mathcal{Z}^{\text{(pole)}} (\tilde{\beta}_{dS},-K) = \exp  \, \frac{\pi \ell^2}{G_N} ~. 
\end{equation}
In the above, the $\mathcal{Z}$ are computed with the bare action, without subtracting the pole on-shell action. The expression parallels a similar observation for two-dimensional near Nariai geometries \cite{Anninos:2022hqo}, and suggests that the thermodynamic content of the empty static patch is purely entropic. It would be interesting to understand the above expression at the one-loop level or higher.

For the black hole patch of pure dS$_4$, we find that, by setting $\frakr_{\text{bh}} \= 0$ in \eqref{eqn: conformal energy entropy bh 4d} and \eqref{eqn: spec heat bh ds4}, the thermodynamic quantities are trivial, $E_\text{conf} \= \mathcal{S}_\text{conf} \= C_K \= 0$, as can be expected.

\subsection{Nariai patch}

In addition to the pure dS$_4$ solution, there is another interesting geometry that can be reached by tuning the conformal temperature $\tilde{\beta}^{-1}$ to a particular value set by the trace of the extrinsic curvature $K\ell$. We call these Nariai patches, which exist as the size of the cosmological and black horizon become close to each other.

To find the Nariai temperature $\tilde{\beta}_N^{-1}$, we consider the following limit. Let $\rho \,\equiv\, \frac{1}{\epsilon}\left(\frakr_\text{tube}-\frac{\ell}{\sqrt{3}}\right)$ be a dimensionless parameter describing deviation of $\frakr_\text{tube}$ from the Nariai radius. The Nariai geometry is obtained by setting $\frakr_{\text{c}} \= \frac{\ell}{\sqrt{3}} + \epsilon$ and taking $\epsilon/\ell \rightarrow 0$ while keeping $\rho$ fixed. Let us first consider the Nariai solution from the cosmic patch perspective. The cosmological horizon and black hole horizons are, respectively, located at $\rho \= 1$ and $-1$. Using \eqref{eqn: cosmic ds4 beta} and \eqref{eqn: cosmo ds4 K}, the Nariai limit fixes
\begin{equation}
	\tilde{\beta}_N \,\equiv\, 2 \pi \sqrt{1-\rho^2} \, , \qquad\qquad K \ell \= \frac{\sqrt{3}\rho}{\sqrt{1-\rho^2}}\, .
\end{equation}
These equations can be inverted analytically to obtain $\tilde{\beta}_N$ in terms of $K\ell$,
\begin{equation}\label{eqn: Nariai temperature}
	\tilde{\beta}_N \= \frac{2\sqrt{3}\pi}{\sqrt{K^2\ell^2+3}} \, .
\end{equation}
We note that $\tilde{\beta}_N$ exists for all values of $K\ell\,\in\, \mathbb{R}$. This is yet another difference with the Dirichlet problem, where to obtain the Nariai solution one needs to fix the value of $\frakr_{\text{tube}}$ to be very close to the Nariai radius.

\textbf{Nariai patch thermodynamics.} Using \eqref{eqn: conformal energy entropy cosmic 4d}, the conformal energy of the Nariai solution is given by
\begin{equation}\label{eqn: rho parameterisation}
	E_\text{conf} \= \frac{\rho\ell^2}{9G_N\sqrt{1-\rho^2}} + \frac{2\ell^2}{3G_N}\sqrt{\frac{\rho^2-4+\sqrt{3}\rho\sqrt{8-5\rho^2}}{9\rho^2-12-\sqrt{3}\rho\sqrt{8-5\rho^2}}} \, .
\end{equation}
The conformal entropy and specific heat, evaluated at constant $K$ and $\tilde{\beta} = \tilde{\beta}_N$, are respectively given by
\begin{equation}
	\mathcal{S}_\text{conf} \= \frac{\pi \ell^2}{3G_N} \, , \qquad\qquad C_K \= - \frac{2\pi\ell^2}{G_N\left(2-9\rho+4\rho^3\right)} \, .
\end{equation}
The specific heat $C_K$ becomes positive (negative) as one pushes the boundary to the cosmological (black hole) horizon. In particular, $C_K$ changes sign at $\rho\=\rho_0$ where\footnote{This is given by the only solution to the cubic equation $4\rho_0^3-9\rho_0+2=0$ with $1>\rho_0>-1$, 
\begin{equation}
    \rho_0 = \sqrt{3}\cos\left(\frac{1}{3}\cos^{-1}\left(-\frac{2}{3\sqrt{3}}\right)-\frac{2\pi}{3}\right)\,.
\end{equation}}
\begin{equation}\label{eqn: K diverge Nariai}
	\rho_0 \,\approx\, 0.227 \qquad \Leftrightarrow \qquad  K \ell \approx 0.405 \,. 
\end{equation}

We can also have black hole solutions in the Nariai patch. These are simply obtained by changing $\rho \to -\rho$ in the expressions above.

The Dirichlet thermodynamics of configurations near the Nariai geometry, from the perspective of a dimensionally reduced theory, was studied in \cite{Svesko:2022txo,Anninos:2022hqo} where it was shown that the cosmic Nariai patch always has negative specific heat.

\subsection{Phase diagram}

\begin{figure}[h!]
        \centering
         \subfigure[]{
                \includegraphics[height=7.5cm]{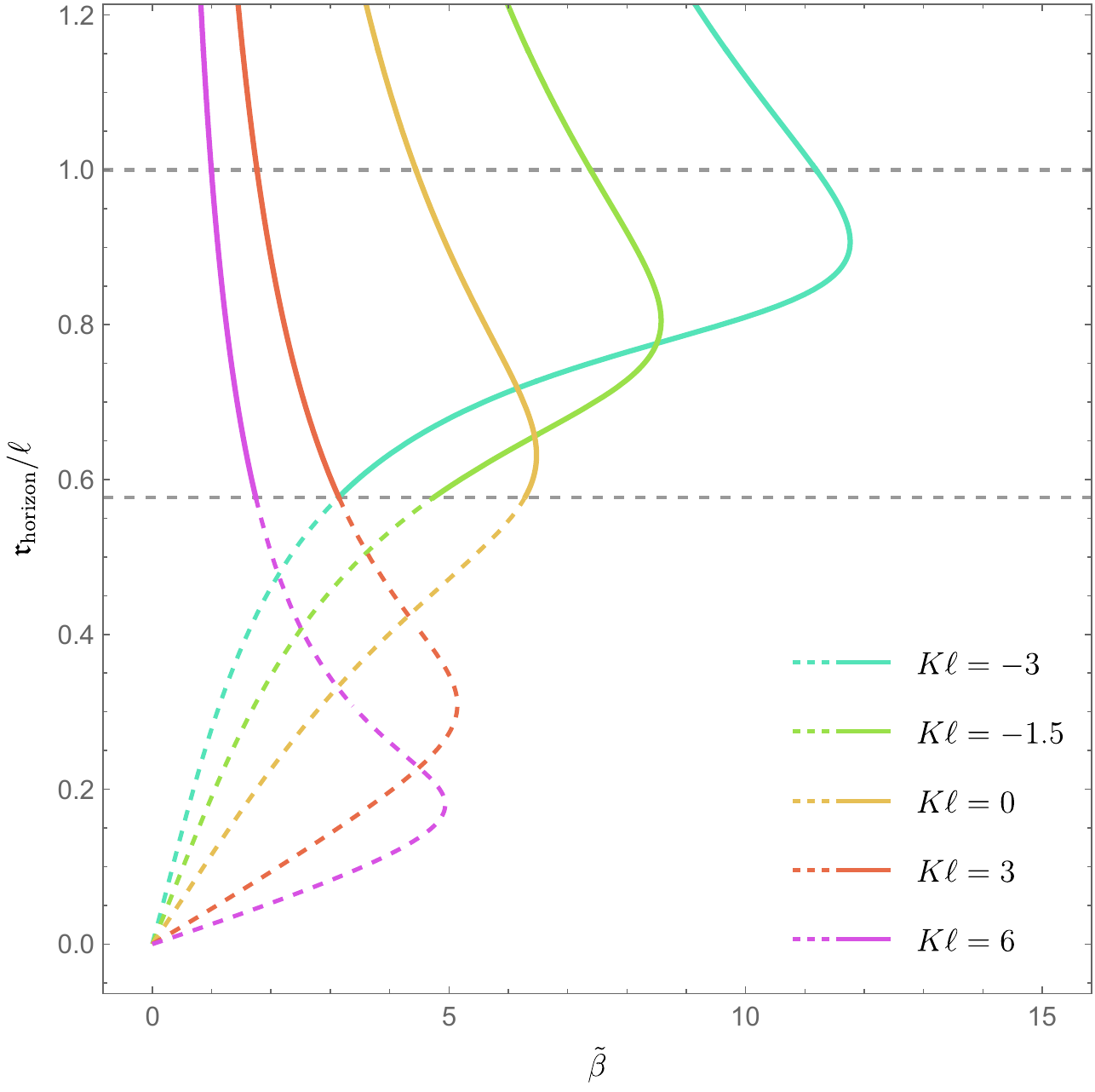}\label{fig: rHorizon 4d}}  \quad\quad
                 \subfigure[]{
                \includegraphics[height=7.5cm]{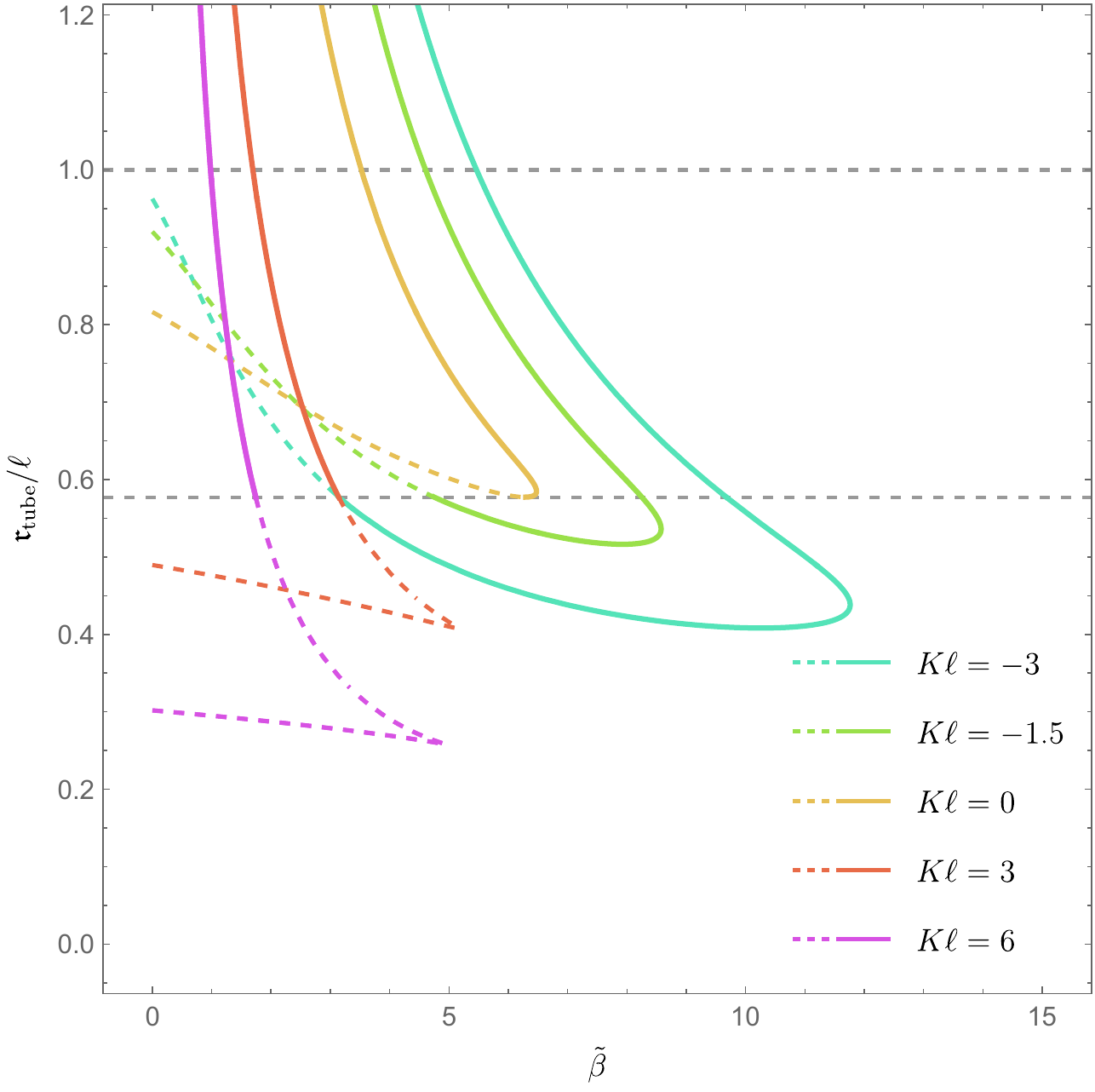} \label{fig: rTube 4d}}  \quad\quad      
                \caption{Plots of $\frakr_\text{horizon}/\ell$ and $\frakr_\text{tube}/\ell$ as a function of $\tilde{\beta}$ at fixed $K\ell$. The solid curves correspond to cosmic patches, while the dashed ones, to black hole patches. $\frakr_\text{horizon}$ denotes the radius of the black hole (cosmological) horizon when considering the black hole (cosmic) patch. The two horizontal black dashed lines correspond to the dS$_4$ patch (upper line) and Nariai patch (lower line).} \label{fig: radius versus beta}
\end{figure}

In this section, we discuss the thermodynamics of four-dimensional spacetime with $\Lambda \,>\,0$ by combining the results from the pole patch, cosmic patch, and black hole patch solutions. In the $G_N \rightarrow 0$ limit, the partition function of the total system is generally given by a sum of all possible patches which have the same boundary data $\tilde{\beta}$ and $K$, 
\begin{equation}\label{eqn: total partition function 4d}
	\mathcal{Z} (\tilde{\beta},K) \= e^{-I^{(\text{pole})}_{E}}\left(1 + \ldots \right) \; .
\end{equation}
A pole patch solution exists for all $\tilde{\beta} \, \in\, \mathbb{R}^+$ and $K\ell \,\in\, \mathbb{R}$. The omitted terms are additional contributions stemming from the co-existing cosmic/black hole patch solutions, i.e. $e^{-I_{E\text{, reg}}^{(\text{cosmic})}}$ or $e^{-I_{E\text{, reg}}^{(\text{black hole})}}$. The number of these terms and the details of the solutions depend on the value of the boundary data, as we will discuss below. For patches with positive specific heat, the one with lowest regulated action is thermodynamically stable; otherwise, they are thermodynamically metastable. Patches with negative specific heat are thermodynamically unstable. 

\begin{figure}[h!]
        \centering
        \includegraphics[scale=0.3]{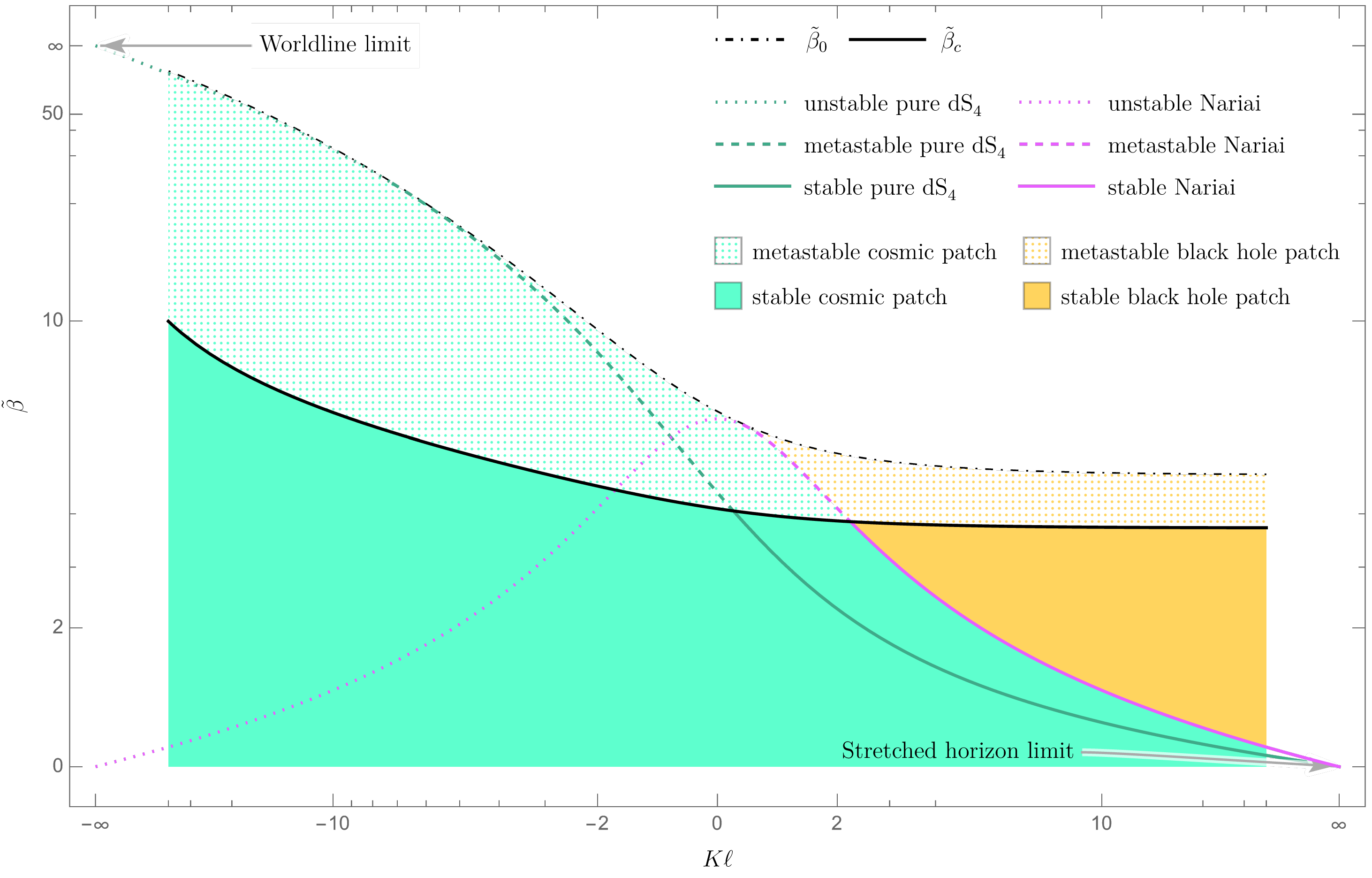}      
                \caption{Phase diagram of conformal dS$_4$ thermodynamics for static and spherically symmetric configurations. The number of different solutions co-existing at a given point in the phase diagram depends on whether the point lies above or below the $\tilde{\beta}_0$ curve (dot-dashed black curve). Above that curve, only one pole patch solution exists. Below the $\tilde{\beta}_0$ curve, apart from a pole patch solution, there co-exist two additional cosmic/black hole patches, one with negative and one with positive $C_K$. The curve of critical inverse conformal temperature $\tilde{\beta}_c$ is shown in thick black, above which the pole patch is thermodynamically preferred. In the region bounded by $\tilde{\beta}_0$ and $\tilde{\beta}_c$ curves, shaded in green (yellow) halftone, the cosmic (black hole) patch is metastable. For $\tilde{\beta}_c > \tilde{\beta}$, the cosmic (black hole) patch is stable with the associated region shaded in solid green (yellow). The dark green and purple curves represent pure dS$_4$ and Nariai patches. Both curves are divided into three segments: stable, metastable, and unstable, which are shown as thick, dashed and dotted curves, respectively. The (meta)stable Nariai curve marks the separation of the (meta)stable cosmic patch and black hole patch regions.} 
                \label{fig: phase4d}
\end{figure}

As opposed to $D=3$, solutions with horizons do not exist for all values of $\tilde{\beta}$ and $K \ell$. At low temperatures, only one pole patch solution exists. To separate the phase space regions with no horizon patches, we define the inverse conformal temperature $\tilde{\beta}_0 (K \ell)$. Note that it depends on the value of $K \ell$, so that at a given $K \ell$, horizon patches only exist for conformal temperatures such that $\tilde{\beta}\leq\tilde{\beta}_0$. The curve $\tilde{\beta}_0 (K \ell)$ can be found numerically and is shown in figure \ref{fig: phase4d}. The rest of the phase diagram can be decribed as follows:

\begin{itemize}
    \item Exactly at $\tilde{\beta}_0 (K \ell)$, there are two solutions: one pole patch and one horizon patch. The latter is a cosmic patch if $K\ell \lesssim 0.405$. Otherwise, it is a black hole patch. Note the transition happens at the $K \ell$ in which the Nariai patch specific heat changes sign. In both cases, the horizon patch has positive regulated action and, therefore, it is always sub-dominant.
    \item For lower inverse temperatures, $\tilde{\beta} < \tilde{\beta}_0 (K \ell)$, we always find three solutions for any given $\tilde{\beta}$ and $K \ell$. There is always one pole patch and two horizon patches. The horizon patches can be either cosmic or black hole patches, but always the horizon patch with larger horizon size has positive specific heat, while the one with smaller size, has negative specific heat. This can be confirmed by observing that the large (small) horizon patch has a horizon radius which is an increasing (decreasing) function of the conformal temperature, as we show in figure \ref{fig: radius versus beta}. Moreover, if the horizon radius is larger (smaller) than the Nariai radius $r_N \= \ell/\sqrt{3}$, the corresponding horizon patch is a cosmic (black hole) patch. There exists a smooth transition between black hole patch and cosmic patch as one varies the conformal temperature.
    \item At a given critical conformal temperature that depends on the value of $K \ell$, there is a first-order phase transition, similar to the Hawking-Page transition. We call this temperature $\tilde{\beta}_c (K \ell)$ and show it numerically in figure \ref{fig: phase4d}. For $\tilde{\beta} < \tilde{\beta}_c$, the large horizon patch solution dominates over the pole patch, while the opposite happens for $\tilde{\beta} > \tilde{\beta}_c$.
\end{itemize}

Consequently, for $\tilde{\beta}_0 > \tilde{\beta} > \tilde{\beta}_c$, the large horizon patch is metastable, while for $\tilde{\beta} < \tilde{\beta}_c$, the large horizon patch becomes stable and the dominant configuration. If a small horizon patch exists, then it is always subdominant.
We display various plots of the regulated action, conformal energy, and specific heat as a function of the inverse conformal temperature at fixed $K\ell$ in appendix \ref{sec: additional figures}, see figure \ref{fig: appendix figures}.

\textbf{Pure dS$_4$ phase structure.} For the pure dS$_4$ solution, we constrain the inverse conformal temperature to be given by the dS inverse temperature \eqref{eqn: dS temp 4d}. We note that $K \ell \,\approx \, 0.256$ is equivalent to $\frakr_\text{tube}/\ell \,\approx\, 0.840$.

\begin{itemize}
    \item For $K\ell\,<\,-7.202$, the dS temperature lies in the intermediate temperature regime implying that the corresponding pure dS$_4$ is sub-dominant. We find that these pure dS$_4$ have negative specific heat and are thus unstable. The worldline limit is included in this regime. At $K\ell \,\approx\, -7.202$, the dS temperature coincides with $\tilde{\beta}_0^{-1}$.
    \item For $-7.202 \, < \, K\ell \, <\, 0.256$, the dS temperature remains in the intermediate regime, but now the associated specific heat becomes positive. Therefore, these pure dS$_4$ patches are metastable. At $K\ell \,\approx\, 0.256$, the dS temperature coincides with the critical temperature, $\tilde{\beta}_c^{-1}$.
    \item For $0.256\,<\,K\ell $, the dS temperature is higher than the critical temperature. As a consequence, the pure dS$_4$ has regulated action lower than the pole patch. It has also positive specific heat. We therefore find that the pure dS$_4$, in this regime, is thermodynamically stable. We note that the strechted horizon is included in this case. 
\end{itemize}

\textbf{Nariai phase structure.} To obtain the Nariai solution, we constrain the inverse conformal temperature to the Nariai inverse temperature \eqref{eqn: Nariai temperature}.
\begin{itemize}
    \item For $K\ell \,<\, 0.405$, the Nariai temperature is higher than $\tilde{\beta}_0^{-1}$ temperature. In this regime, the Nariai solution has negative specific heat, so it is thermally unstable.
    \item For $0.405\,<\,K\ell\,<\,2.239$, the Nariai temperature lies in the intermediate temperature regime. The corresponding Nariai solution has positive specific heat and positive regulated action. This means that the Nariai soution, in this regime, is metastable.
    \item For $2.239\,<\,K\ell$, the Nariai temperature is higher than the critical temperature. We find that the Nariai solution now becomes thermodynamically stable.
\end{itemize}


\section{Linearised dynamics} \label{sec: dynamics}

So far our treatment has been largely based on a quasi-equilibrium Euclidean picture. The aim of our final section is to complement the Euclidean analysis with  a Lorentzian analysis. Concretely, we will consider solutions to the four-dimensional linearised Einstein equations equipped with a positive cosmological constant $\Lambda\,>\,0$. As boundary conditions, we will once again consider conformal boundary conditions for the induced metric $g_{mn}$ and mean curvature $K$ on a topologically $\mathbb{R} \times S^2$ timelike boundary $\Gamma$. Our treatment parallels that for Minkowski space \cite{Anninos:2023epi}, here extended to the case of $\Lambda\,>\,0$. A portion of our linearised analysis was already treated in \cite{Anninos:2011zn}, in the context of the fluid-gravity correspondence applied to de Sitter horizons.\footnote{The conformal boundary conditions were necessitated in \cite{Anninos:2011zn}, as well as \cite{Bredberg:2011xw}, due to an obstruction in solving the non-linear Einstein equations with Dirichlet data on a timelike surface in the near-horizon expansion. The Lorentzian problem with Dirichlet conditions was considered in \cite{Andrade:2015gja}, where exponentially growing modes were found for the cosmic patch, as well as the pole patch at sufficiently large worldtube size.}

\subsection{Basic setup}


We will consider the linearised Einstein equations about the static patch metric,
\begin{equation}\label{eqn: lorentzian background metric}
	ds^2 \= -f(r)dt^2 + \frac{dr^2}{f(r)} + r^2 d\Omega^2 \, , \qquad f(r) \= 1-\frac{r^2}{\ell^2} \, , \qquad d\Omega^2 \= d\theta^2 + \sin^2\theta d\phi^2 \, .
\end{equation}
The timelike boundary $\Gamma$ is located at $r=\frakr \in \left(0,\ell\right)$. As in \cite{Anninos:2011zn}, we are primarily interested in dynamical features of the cosmological horizon, but we also report on the dynamical features of the pole patch below. As such, the spacetime region of interest is taken to be the Lorentzian cosmological patch $r \in \left(\frakr,\ell\right)$, $t \in \mathbb{R}$, and $\theta \in \left(0,\pi\right)$, $\phi \sim \phi + 2\pi$. The induced metric on $\Gamma$ is given by
\begin{equation}\label{eqn: lorentzian induced metric}
	\left.ds^2 \right|_{r=\frakr} \= -f(\frakr) dt^2 + \frakr^2 d\Omega^2  \, .
\end{equation}
Using an inward-pointing normal vector $\hat{n}\=-\sqrt{f(r)}\partial_r$, the extrinsic curvature and its trace are given by,
\begin{equation}\label{eqn: lorentzian extrinsic}
	\left.K_{mn} dx^m dx^n\right|_{r=\frakr} \= \frac{\frakr\sqrt{f(\frakr)}}{\ell^2}\left(dt^2 + \ell^2 d\Omega^2\right)~, \quad\quad \left.K\ell\right|_{r=\frakr} \= \frac{3\frakr^2-2\ell^2}{\frakr\sqrt{\ell^2-\frakr^2}} \, .
\end{equation}
We denote linearised perturbations about the background (\ref{eqn: lorentzian background metric}) as
\begin{equation}
	g_{\mu\nu} \= \bar{g}_{\mu\nu} + \varepsilon \, h_{\mu\nu} \, , \qquad\qquad \left|\,\varepsilon\,\right| \ll 1 \, ,
\end{equation}
where the background metric $\bar{g}_{\mu\nu}$ is given in \eqref{eqn: lorentzian background metric}. The equation of motion for $h_{\mu\nu}$ is obtained by expanding \eqref{eqn: Einstein field} to first order in $\varepsilon$. Further demanding that the conformal boundary data remains invariant under arbitrary perturbation $h_{\mu\nu}$ implies that
\begin{eqnarray}\label{eqn: def bdry cond 1}
\begin{cases}
	\left.h_{mn}\right|_{r=\frakr} &= \left.\gamma(x)\bar{g}_{mn}\right|_{r=\frakr} \, , \\
	\left.\varepsilon \delta K(h_{\mu\nu})\right|_{r=\frakr} & \equiv \left.K(\bar{g}_{\mu\nu}+\varepsilon \, h_{\mu\nu}) - K(\bar{g}_{\mu \nu})\right|_{r=\frakr} \= 0 \, ,
	\end{cases}
\end{eqnarray}
where $\gamma(x)$ is an arbitrary function, which will depend on the initial data of the linearised metric $h_{\mu\nu}$, and $\bar{g}_{mn}|_{r=\frakr}$ is the induced metric \eqref{eqn: lorentzian induced metric}. By contracting the first expression in \eqref{eqn: def bdry cond 1} with $\bar{g}^{mn}$, one may write the first boundary condition in a form that does not contain $\gamma(x)$ as
\begin{equation}\label{eqn: bdry cond 1}
	\left.h_{mn} - \frac{1}{3}\bar{g}_{mn} h^p{}_p \right|_{r=\frakr} \= 0 \, .
\end{equation}
Using \eqref{eqn: def of K}, the variation of the trace of the extrinsic curvature to first order in $\epsilon$ is given by
\begin{equation}\label{eqn: bdry cond 2}
	\left.\frac{\delta K (h_{\mu\nu})}{\sqrt{f(\frakr)}}\right|_{r=\frakr} \= \left.\frac{1}{2} \partial_r h^m{}_m - \mathcal{D}^m h_{rm} - \frac{\sqrt{f(\frakr)}}{2}K h_{rr} \right|_{r=\frakr} = 0 \, ,
\end{equation}
where $\mathcal{D}_n$ denotes the covariant derivative with respect to the boundary metric $\bar{g}_{mn}$. In the following analysis, we will take \eqref{eqn: bdry cond 1} and \eqref{eqn: bdry cond 2} as the conformal boundary conditions for linearised gravity with $\Lambda>0$. We must also impose conformal boundary conditions on the space of allowed diffeomorphisms $\xi_\mu$. Finally, we  require that $\xi^r |_{r=\frak{r}} = 0$, such that the allowed diffeomorphisms do not move the location of the boundary.

{\textbf{Kodama-Ishibashi method.}} Following the treatment of Kodama and Ishibashi \cite{Kodama:2000fa,Kodama:2003jz}, due to the spherical symmetry and time-translation invariance of the background, we can split our linearised solutions into vector and scalar perturbations, denoted by $h^{(V)}_{\mu\nu}$ and $h^{(S)}_{\mu\nu}$ respectively. Our details and conventions follow directly those in appendix C of \cite{Anninos:2023epi}. As such, our treatment will be brief in what follows and mostly focused on presenting the main results for $\Lambda>0$.

The $SO(3)$ content of $h^{(V)}_{\mu\nu}$ is captured by the vectorial spherical harmonics, $\mathbb{V}_i$, which are transverse eigenfunctions of the unit two-sphere Laplacian acting on vectors, with eigenvalues $k_V = l(l+1)-1$ for $l = 1,2,\ldots$ The $SO(3)$ content of $h^{(S)}_{\mu\nu}$ is captured by the scalar spherical harmonics, $\mathbb{S}$, which are transverse eigenfunctions of the unit two-sphere Laplacian with eigenvalues $k_S = l(l+1)$ for $l = 0,1,2,\ldots$  We note that the $l=0$ and $l=1$ modes require a separate treatment and we discuss them in appendix \ref{sec: l0/l1 modes}. Together, $h^{(V)}_{\mu\nu}$ and $h^{(S)}_{\mu\nu}$ encode the two propagating degrees of freedom of the four-dimensional metric at the linearised level. 

In the absence of timelike boundaries, the Kodama-Ishibashi formalism is gauge invariant and reduces the linearised Einstein equations to a set of `master equations' governing the vectorial and scalar master fields $\Phi^{(V)}$ and $\Phi^{(S)}$ which are directly linked to $h^{(V)}_{\mu\nu}$ and $h^{(S)}_{\mu\nu}$. It proves convenient for our analysis, as it did in \cite{Anninos:2023epi}, to select a gauge where the linearised boundary conditions (\ref{eqn: bdry cond 1}) and (\ref{eqn: bdry cond 2}) act only on $h^{(V)}_{\mu\nu}$ and $h^{(S)}_{\mu\nu}$ respectively. This gauge choice is indeed possible, and in this gauge the components of our metric perturbation read
\begin{eqnarray}
\begin{cases}
	h_{mn} &= - \bar{g}_{mn} \frac{1}{2r}\left[ l\left(l+1\right)\left(1- \frac{2r^2}{\ell^2}\right)+2 r^2 \partial_t^2 + 2  \left(1-\frac{r^2}{\ell^2}\right)^2  r \partial_r\right]\Phi^{(S)} \mathbb{S}  \\
 & \quad + \left(\delta^i_m\delta^t_n + \delta^i_n \delta^t_m\right) \left(1-\frac{r^2}{\ell^2}\right)\partial_r \left(r \Phi^{(V)}\right) \mathbb{V}_i \, , \\
    h_{rr} &= -\frac{1}{r\left(1-\frac{r^2}{\ell^2}\right)^2}\Bigg[\frac{l(l+1)}{2}\left(3-\frac{7r^2}{\ell^2}+\frac{4r^4}{\ell^4}\right)+ \left(3-\frac{2 r^2}{\ell^2}\right)r^2\partial_t^2 \\
    		&\quad + \left(1-\frac{r^2}{\ell^2}\right)\left(\left(1-\frac{r^2}{\ell^2}\right)\left(l(l+1) + 1-2\frac{r^2}{\ell^2} \right)+ r^2\partial_t^2\right) r\partial_r\Bigg]\Phi^{(S)} \mathbb{S} \, , \\
    h_{tr} &=  -\frac{1}{2\left(1-\frac{r^2}{\ell^2}\right)}\partial_t \left[l(l+1)\left(1-\frac{r^2}{\ell^2}\right) - 2 + r^2\partial_t^2 - \left(1-\frac{r^2}{\ell^2}\right)\frac{r^2}{\ell^2}r\partial_r\right]\Phi^{(S)} \mathbb{S} \, ,  \\
    h_{ri} &=  \frac{\sqrt{l(l+1)}}{2\left(1-\frac{r^2}{\ell^2}\right)}\left[l(l+1)\left(1-\frac{r^2}{\ell^2}\right)+r^2\partial_t^2+\left(2-\left(3-\frac{r^2}{\ell^2}\right)\frac{r^2}{\ell^2}\right)r\partial_r\right]\Phi^{(S)} \mathbb{S}_i + \frac{r}{1-\frac{r^2}{\ell^2}} \partial_t \Phi^{(V)} \mathbb{V}_i \, .	
    \end{cases}\label{eqn: spherical l>2 ansatz}
\end{eqnarray}
In the above the indices $m$ and $n$ denote indices with respect to $(t,r)$, while the index $i$ denotes indices on the two-sphere.

\subsection{Vector perturbation}

The master equation for $\Phi^{(V)}$, for given angular momentum $l \in \mathbb{Z}^+$, is given by
\begin{equation}
\left(-\boldsymbol{\nabla}^2  + \frac{l(l+1)}{r^2} \right) \Phi^{(V)}(t,r) = 0~,
\end{equation}
where $\boldsymbol{\nabla}^2$ denotes the Laplacian on a two-dimensional de Sitter space with curvature $+2/\ell^2$. The solutions can be expressed as hypergeometric functions  (see for instance \cite{Bredberg:2011xw}). For a given frequency, they take the form
\begin{equation}
    \Phi^{(V)} \= \Re\, e^{-i\omega t}\left(1-\frac{r^2}{\ell^2}\right)^{-i\omega\ell/2} \left(\frac{r^2}{\ell^2}\right)^{i\omega\ell/2} {}_2F_1\left(-l-i\omega \ell,1+l-i\omega\ell;1-i\omega\ell;\frac{1}{2}-\frac{\ell}{2r}\right) \, .
\end{equation}
The boundary condition \eqref{eqn: bdry cond 2} is automatically satisfied while the boundary condition \eqref{eqn: bdry cond 1} imposes
\begin{equation}
    \left.\frac{\Phi^{(V)}}{r} + \partial_r \Phi^{(V)}\right|_{r=\frakr}\=0 \, .
\end{equation}
Upon scanning numerically for solutions in the complex frequency plane, we find that all vector modes satisfying the conformal boundary conditions have a negative imaginary part, and are therefore dissipative, decaying at late times. 

\textbf{Worldline limit.} In the worldline limit, where $K\ell \to -\infty$, we find two sets of modes. One set is found to be a small deformation of the quasinormal modes of the de Sitter static patch, whose analytic form (in the worldline limit) is given by \cite{Lopez-Ortega:2006aal} $\omega^{\text{qnm}} \ell = - i (l+n+1)$ where $n\in\mathbb{N}$. As for the exact quasinormal modes, the modes we find are also purely negative imaginary and their size is of the order of the de Sitter length $\ell$. The other set of modes have a real part also and are the counterpart of the vectorial Minkowski modes uncovered in \cite{Anninos:2023epi}. For each $l \ge 2$, the second set is a discrete tower of modes with increasing negative imaginary parts, and their size scales with the size of the worldtube $\frak{r}$, rather than the de Sitter length. 

\textbf{Cosmological horizon limit.} In the cosmological horizon limit, where $K\ell \to +\infty$, the structure of the modes is altered. The purely imaginary modes degenerate into a set of modes that approach 
\begin{equation}\label{shear}
    \omega^{\text{shear}} \ell  \= - i \left(l(l+1)-2\right) \frac{1}{2 K^2\ell^2} + \mathcal{O}(K^{-3}\ell^{-3}) \, .
\end{equation}
These modes were identified in \cite{Anninos:2011zn} where they were interpreted as shear modes of a linearised incompressible non-relativistic fluid dynamical behavior near the horizon, paralleling other considerations of the fluid/gravity relation \cite{Damour:1978cg,znajek1978electric,Price:1986yy,Bhattacharyya:2008kq,Bredberg:2011xw}. For each $l \ge 2$, the second set is a discrete tower of modes with increasing negative imaginary parts, and their size scales with the de Sitter length.

\subsection{Scalar perturbation}\label{sec: scalar pert}

The master equation for $\Phi^{(S)}$ is given by
\begin{equation}
\left(-\boldsymbol{\nabla}^2  + \frac{l(l+1)}{r^2} \right) \Phi^{(S)} = 0~.
\end{equation}
The solutions can be expressed as hypergeometric functions, and take the form
\begin{equation}\label{PhiS}
    \Phi^{(S)} \= \Re\, e^{-i\omega t}\left(1-\frac{r^2}{\ell^2}\right)^{-i\omega\ell/2} \left(\frac{r^2}{\ell^2}\right)^{i\omega\ell/2} {}_2F_1\left(-l-i\omega \ell,1+l-i\omega\ell;1-i\omega\ell;\frac{1}{2}-\frac{\ell}{2r}\right) \, .
\end{equation}
The boundary condition \eqref{eqn: bdry cond 1} is automatically satisfied while the boundary condition \eqref{eqn: bdry cond 2} imposes
\begin{eqnarray}\label{eqn: bdry cond scalar perturbation}
	\mathcal{F}_l(K\ell\,,  \omega\ell) \,\equiv\,\left.\left(\frac{a_1}{r^4}+\frac{a_2}{r^2}\partial_t^2 - 2 \partial_t^4\right)\Phi^S  + \left(\frac{a_3}{r^2} -2 \partial_t^2\right)\left(1-\frac{r^2}{\ell^2}\right)^2\frac{\partial_r \Phi^{(S)}}{r} \right|_{r=\frakr} \= 0 \, ,
 \end{eqnarray}
 where
 \begin{eqnarray}
     \begin{cases}
	& a_1  \= l(l+1)\left(1-\frac{r^2}{\ell^2}\right)\left(3-\frac{2r^2}{\ell^2}-2l(l+1)\left(1-\frac{r^2}{\ell^2}\right)\right) \, , \\
	& a_2  \= 4-\frac{2r^2}{\ell^2} - 4 l(l+1)\left(1-\frac{r^2}{\ell^2}\right) \, , \\
	& a_3  \= 4-\frac{2r^2}{\ell^2}-l(l+1)\left(3-\frac{2r^2}{\ell^2}\right) \, .
    \end{cases}
\end{eqnarray}
Upon scanning numerically for solutions in the complex frequency plane, we find that some scalar modes satisfying the conformal boundary conditions have a positive imaginary part. 

\textbf{Worldline limit.} In the worldline limit, where $K\ell \to -\infty$, we find two sets of allowed frequencies. One set corresponds to a small deformation of the scalar quasinormal modes in the pure static patch, which are of the order of the de Sitter length scale $\ell$. The deformed quasinormal mode frequencies have a negative imaginary part, and are hence dissipative. The other set of allowed frequencies is borrowed from the analogous modes in Minwkoski space uncovered in \cite{Anninos:2023epi} and are of the order of the worldtube size $\mathfrak{r} \ll \ell$. For this set of modes, we find a pair of allowed frequencies with positive imaginary part for each $l$. 

\textbf{Cosmological horizon limit.}  In the cosmological horizon limit, where $K\ell \to +\infty$, the structure of the modes is altered. There is still a collection of fluid-type modes, but they take a relativistic dispersion relation. Upon expanding $r$ near $\ell$ in (\ref{PhiS}), implementing the conformal boundary conditions, and scaling $\omega$ with $1/K$, we find the analytic expansion
\begin{equation}
    \omega^{\text{sound}} \ell \= \pm  \frac{1}{\sqrt{2}K\ell} \sqrt{l(l+1)} - i\frac{l(l+1)-2}{2}  \frac{1}{2K^2\ell^2}+ \mathcal{O}(K^{-3}\ell^{-3})\, .
\end{equation}
It is natural to interpret the above modes as the sound mode counterpart to the fluid dynamical shear modes in (\ref{shear}). It is worth noting, however, that they scale differently with $K\ell$, such that in the strict horizon limit only the shear modes survive. This is one of the reasons the sound modes, whose speed of sound becomes infinite in this limit, does not appear in the previous analyses of \cite{Bredberg:2011xw,Anninos:2011zn}. 

\begin{figure}[h!]
        \centering
         \subfigure[$l = 2$]{
                \includegraphics[width = 7.51 cm]{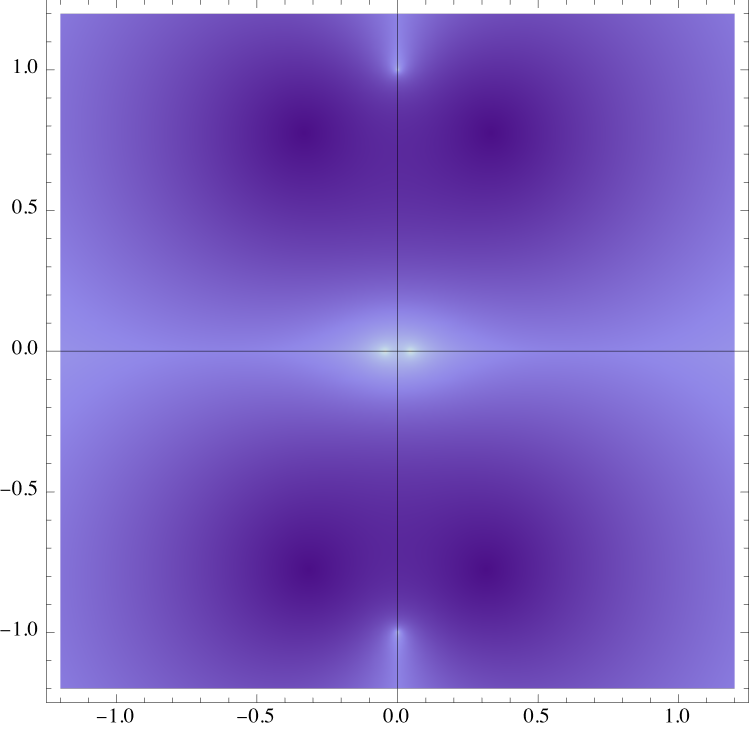}\label{fig: l2}}  \quad\quad
                 \subfigure[$l = 10$]{
                \includegraphics[width= 7.51 cm]{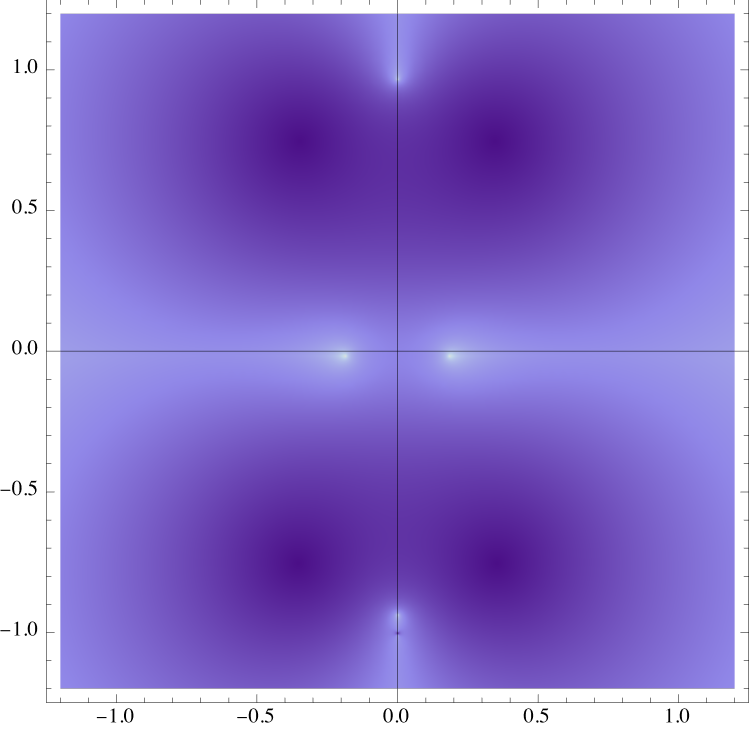} \label{fig: l10}}                           
                \caption{Density plot of absolute value of $\log \left(e^{-4 
                \omega \ell i}\mathcal{F}_l(K\ell\,,  \omega\ell)\right)$ in the complex $\omega\ell$ plane for $l=2$ and $l=10$, where $\mathcal{F}_l(K\ell\,,  \omega\ell)$ is defined in \eqref{eqn: bdry cond scalar perturbation}. In both plots, $K\ell$ is fixed to be $40$. Both the $\omega\ell \approx \pm i$ and $\omega^\text{sound}\ell$ are displayed in both plots. For $l=10$, the sound modes develop a small negative imaginary part.} \label{fig: omega density plot}
\end{figure}

In addition to the sound modes, upon taking the strict horizon limit, $K\ell \to +\infty$, of the scalar solutions (\ref{PhiS}) for each $l$,  and implementing the conformal boundary conditions, any solutions with modes with positive imaginary frequency coalesce either to $\omega \ell \= + i$ or $\omega \ell \= - i$. We show these in figure \ref{fig: omega density plot}. One can identify these modes in a Rindler analysis, subject to conformal boundary conditions. Concretely, we take the Rindler metric to be
\begin{equation}
    ds^2 \= -\frac{z^2}{z_0^2} dt^2 + dz^2 + dx^2 + dy^2~,
\end{equation}
with $(t, x,y) \,\in\, \mathbb{R}^3$ and $z \,\in\, \left(0,z_0\right]$. We then perform a straightforward analysis of the linearised Einstein equations, with vanishing $\Lambda$, subject to conformal boundary conditions at $z=z_0$. One observes that the following configuration (chosen for simplicity to have spatial momentum entirely along the $x$-direction)
\begin{equation}\label{rindler}
    \begin{cases}
        h_{tt} &=\, - \frac{z^2}{z_0^2} h_{xx} \= - \frac{z^2}{z_0^2} h_{yy} \= -\frac{z^2}{z_0^2}e^{-i\left(\omega t - k_x x\right)} J_{-i\omega z_0}(-ikz) \, , \\
        h_{zz} &=\, \frac{e^{-i\left(\omega t - k_x x\right)}}{k^2z^2} \left(\left(k^2z^2 + 2 \omega^2 z_0^2\right)  + \left(2k^2z^2-2\omega^2z_0^2\right)z\partial_z\right) J_{-i\omega z_0}(-ikz) \, , \\
        h_{zt} &=\, \frac{i\omega e^{-i\left(\omega t - k_x x\right)}}{k^2 z}\left(\left(2-k^2z^2+\omega^2z_0^2\right) - z \partial_z\right) J_{-i\omega z_0}(-ikz) \, , \\
        h_{zx} &=\, -\frac{ie^{-i\left(\omega t - k_x x\right)}}{k_x z} \left(\left(-k^2z^2 + \omega^2z_0^2\right) + z \partial_z\right)  J_{-i\omega z_0}(-ikz)\, ,
    \end{cases}
\end{equation}
solves the linearised Einstein equations with $\Lambda \= 0$, subject to conformal boundary conditions at $z=z_0$, for a selection of complex frequencies. In the limit $k z_0 \to 0$ the allowed frequencies coalesce to $\omega z_0 \= \pm i$. This mode coincides with the linearised de Sitter mode with $\omega\ell= \pm i$. Upon expressing the Rindler solution (\ref{rindler}) in terms of a local inertial time coordinate, one notes that it grows at most polynomially. As such, although growing in time, the exponential growth of (\ref{rindler}) is more tame than an ordinary unstable mode. In fact, to leading order at small $k z_0$, the Rindler mode is locally a pure diffeomorphism. 

As shown in appendix \ref{diffeoapp}, the leading contribution to the de Sitter scalar modes (\ref{PhiS}) with $\omega = \pm i\ell$, in the stretched cosmological horizon limit, are locally pure gauge. This leads to a double suppression effect whereby the physical contribution of the modes is not only small due to the linearised nature of $h_{\mu\nu}$, but also due to a suppression factor that goes as $1/K^2\ell^2$.

\subsection{Dynamics of the pole patch, briefly}

One can also consider linearised gravity subject to conformal boundary conditions in the pole patch. Here we further impose that the gravitational solutions are  smooth throughout the whole interior. The pole patch has been explored as a potential candidate for static patch holography in \cite{banks2012holographic,Coleman:2020jte,Susskind:2021esx,Shaghoulian:2021cef,Jorstad:2022mls} among other places. 

\textbf{Worldline limit.} For $K\ell \to +\infty$ the size of the timelike boundary is small in units of the de Sitter length $\ell$. Here our analysis matches the Minkowski analysis presented in \cite{Anninos:2023epi}, where it was observed that the allowed vector modes frequencies are all real-valued, whilst the scalar mode frequencies permit a subset of modes $\omega^{(S)}$ with positive real imaginary frequency for each $l$. At large $l$, these modes were numerically found to scale as $\omega^{(S)} \frak{r} \approx \pm l + i c l^{1/3}$, where $c$ is an order one number. Thus, the pole patch in the thin world tube limit mimics the Minkowskian picture. 


\textbf{Cosmological horizon limit.} For $K\ell \to -\infty$ the timelike boundary of the pole patch approaches the cosmological horizon. In this regime, the analysis differs from the thin worldline limit. Nonetheless, we observe the presence of scalar modes with complex frequencies of positive imaginary part for each $l$. These modes, similarly to the cosmic patch,  coalesce onto $\omega^{\text{pole}}\ell \= \pm i$. Moreover, we find two additional sets of modes. The first set is a pair of real valued scalar modes for each $l$. Numerically, at large $l$, they are found to be proportional to $\tfrac{l}{K\ell}$. The second set is an infinite tower of real valued modes, appearing both in the vector and scalar sector. At large $l$, they are numerically found to be evenly spaced and their magnitude is inversely proportional to $\log|\sqrt{2}K\ell|$.

\begin{center}
\pgfornament[height=5pt, color=black]{83}
\end{center}
\vspace{5pt}

We can synthesise, in short. We have provided evidence that, subjected to conformal boundary conditions, the stretched horizon limit of the pure de Sitter patch is a thermodynamically stable portion of spacetime containing a cosmological horizon. Dynamically, the majority of linearised gravitational perturbations about this portion of spacetime decay at late times, save one mode for each $l$ of total angular momentum. These modes have a purely imaginary frequency $\omega\ell = + i$, and are moreover retrieved from a Rindler analysis. We take this latter property as an indication that they are not endemic to de Sitter space, but rather a universal property of near horizon physics subjected to conformal boundary conditions. Their fate, whose behavior as measured by a local inertial clock is at most polynomial in time, is a remaining obstacle in obtaining a portion of spacetime with $\Lambda>0$ that is both thermodynamically stable, as well as dynamically stable at the linearised level. Perhaps to tame this mode we must impose an additional boundary condition, always ensuring that in doing so we do not overly restrict any interesting dynamics. Or perhaps we must relax the condition of a constant $K$. A careful examination is left to the future. 




\section*{Acknowledgements}
It is a pleasure to acknowledge Tarek Anous, Eleanor Harris, Diego Hofman, Juan Maldacena, Beatrix M\"uhlmann, Edgar Shaghoulian, Eva Silverstein and Manus Visser for interesting discussions. We are particularly grateful for interesting discussions with Andrew Svesko. D.A. is funded by the Royal Society under the grant ``Concrete Calculables in Quantum de Sitter". The work of D.A.G. is funded by UKRI Stephen Hawking Fellowship ``Quantum Emergence of an Expanding Universe". D.A. and D.A.G. are further funded by STFC Consolidated grant ST/X000753/1. C.M. is funded by STFC under grant number ST/X508470/1.

\appendix

\section{dS$_3$ Dirichlet thermodynamics} \label{sec:3d_Dirichlet}
In this appendix, we review the thermodynamics of the three-dimensional dS using Dirichlet boundary condition \cite{Coleman:2021nor}. Here, we use the standard coordinate for Euclidean Schwarzschild-de-Sitter space
\begin{equation}
	ds^2 \= \frac{f(r)}{f(\frakr_\text{tube})}d\tau^2 + \frac{dr^2}{f(r)} + r^2 d\phi^2 \, , \qquad f(r) \= \frac{\frakr_{\text{c}}^2-r^2}{\ell^2} \, ,
\end{equation}
where $\tau \,\sim\, \tau + \beta_D$ and $\phi \,\sim\, \phi + 2\pi$. The Euclidean time $\tau$ is measured with respect to the clock defined on the boundary $r\=\frakr_\text{tube}$. Similar to the analysis in the main text, $\frakr_c$ represents the physical size of the cosmological horizon. There is no black hole in this geometry. When $\frakr_c \neq 1$, there is a conical defect at the origin $r\=0$.

The boundary $r\=\frakr_\text{tube}$ has topology of $S^1 \times S^1$ and obeys Dirichlet boundary data, namely the induced metric at $r\=\frakr_\text{tube}$ is fixed to be
\begin{equation}
	\left.ds^2\right|_{r=\frakr_\text{tube}} \= d\tau^2 + \frakr_\text{tube}^2 d\phi^2 \, .
\end{equation} 
In this case, the boundary data consists of the Euclidean time period $\beta_D$ and the physical size of the boundary $\frakr_\text{tube}$. The Euclidean action $I_E$ is given by \eqref{euclidean_action} with $\alpha_{\text{b.c.}} = (D-1)$ and $D=3$.


Since the Dirichlet problem fixes $\beta_D$ and $\frakr_\text{tube}$, the partition function is a function of these, $\mathcal{Z}\=\mathcal{Z}(\beta_D,\frakr_\text{tube})$. The energy $E$ and entropy $\mathcal{S}$ follow from \eqref{eqn: defining thermo quant} with $\beta_D$ replacing $\tilde{\beta}$. The specific heat now becomes the specific heat at constant $\frakr$, $C_\frakr$, and is defined as in \eqref{eqn: defining thermo quant}, but with $\frakr_\text{tube}$ replacing $K$.

The solutions are divided into two classes as in the conformal problem, the pole patch and cosmic patch.

\textbf{Pole patch.} The first class of solutions is the pole patch solutions which restrict the spacetime to $r\,\in\,\left[0,\frakr_\text{tube}\right]$. In order to have a regular geometry, we must set $\frakr_c\=\ell$. Consequently, $\beta_D$ is a free parameter, and the Euclidean action is given by
\begin{equation}\label{eqn: Dirichlet euclidean action 1 3d}
	I_E^{(\text{pole})} \= -\frac{\beta_D \frakr_\text{tube}}{4G_N\ell}\sqrt{\frac{\ell^2}{\frakr_\text{tube}^2}-1} \, .
\end{equation}
It follows that the energy is given by
\begin{equation}
	\beta_D E \= {I_E^{(\text{pole})}} \, ,
\end{equation}
and the entropy and specific heat at constant $\frakr_\text{tube}$ are zero.

\textbf{Cosmic patch.} We now consider solutions with $r\,\in\, \left[\frakr_\text{tube},\frakr_{\text{c}}\right]$, so they contain the cosmological horizon. Regularity near the horizon $\frakr_c$ fixes
\begin{equation}
	\beta_D \= 2\pi\ell \sqrt{1- \frac{\frakr_\text{tube}^2}{\frakr_{\text{c}}^2}} \, .
\end{equation}
The Euclidean action for this case is given by
\begin{equation}\label{eqn: Dirichlet euclidean action 2 3d}
	I_E^{(\text{cosmic})} \= -\frac{\beta_D \frakr_\text{tube}}{4G_N \ell}\sqrt{\left(1-\frac{\frakr_\text{tube}^2}{\frakr_{\text{c}}^2}\right)^{-1}-1} \, .
\end{equation}
We note that, by setting $\frakr_{\text{c}} \=\ell$, \eqref{eqn: Dirichlet euclidean action 1 3d} and \eqref{eqn: Dirichlet euclidean action 2 3d} reproduce (B.8) and (B.16), respectively, in \cite{Coleman:2021nor}, but without adding the additional boundary subtraction term.

The energy of the cosmic patch is given by
\begin{equation}
	E \= \frac{1}{4G_N\ell}\sqrt{\frakr_{\text{c}}^2-\frakr_\text{tube}^2} \, .
\end{equation}
Defining a dimensionless energy $\mathcal{E} \,\equiv\, 2 \pi E \, \frakr_\text{tube} $, we find that
\begin{equation}
	\mathcal{E} + \frac{\pi \frakr_\text{tube}^2}{2 G_N \ell} \= \frac{\pi \frakr_\text{tube}^2}{2G_N \ell}\left(1+ \sqrt{-1+\frac{\frakr_{\text{c}}^2}{\frakr_\text{tube}^2}}\right) \, , 
\end{equation}
which agrees with the energy in \cite{Coleman:2021nor}, upon identifying
\begin{equation}
	\frac{3\ell}{2G_N} \to c  \, , \qquad\qquad \frac{\frakr_{\text{c}}^2}{\ell^2} \to \frac{12 \Delta}{c} - 1 \, , \qquad\qquad \frac{2G_N \ell}{\pi^2 \frakr_\text{tube}^2}  \to y \, .
\end{equation}

We can also compute the entropy and the specific heat with Dirichlet boundary conditions. For the entropy we obtain,
\begin{equation}
\mathcal{S} \= \frac{\pi \frakr_{\text{c}}}{2G_N} \,,
\end{equation}
which agrees with the Gibbons-Hawking entropy. In fact, this entropy (including a sub-leading logarithmic correction) can be matched to the entropy in a $T\bar{T} + \Lambda_2$-deformed CFT$_2$ \cite{Coleman:2021nor}. Finally, for the specific heat we obtain,
\begin{equation}
C_{\text{(Dirichlet)}} \= - \frac{\pi \frakr_{\text{c}}}{2G_N}\left(\frac{\frakr_{\text{c}}^2}{\frakr_\text{tube}^2}-1\right) \,,
\end{equation}
which is negative for all values of $\mathfrak{r}_{\text{tube}}$. Moreover, in the worldline limit, the specific heat diverges.

\section{Useful formulae for $D=4$}\label{sec: useful formulae for 4d thermo}
In this appendix, we provide useful formulae to study conformal thermodynamics of four-dimensional spacetime with $\Lambda \,>\, 0$. 

First, we recall the metric \eqref{eqn: euclidean sol 4D},
\begin{equation}
	ds^2 \= e^{2\omega} \left(\frac{f(r)}{f(\frakr)}d\tau^2 + \frac{dr^2}{f(r)} + r^2 d\theta^2 + r^2 \sin^2{\theta} d\phi^2 \right)\, , \qquad f(r) \= 1- \frac{2\mu}{r} - e^{2\omega} \frac{r^2}{\ell^2}\, .
\end{equation} 
Using a normal vector $\hat{n}\= \sqrt{f(r)}\partial_r$, one can express $e^{2\omega}$ in terms of $K\ell$ and $\mu/\frakr$. For $0\,<\,\mu/\frakr\,<\,1/3$, there exists a unique $e^\omega$ for any real $K \ell$ given by 
\begin{equation}\label{eqn: appendix dum1}
    e^{2\omega} \= \frac{3+\left(K^2\ell^2+9\right) \left(1-\frac{2\mu}{\frakr}\right) - K\ell \sqrt{\left(K^2\ell^2+9\right)\left(1-\frac{2\mu}{\frakr}\right)^2-1}}{2 \left(K^2\ell^2+9\right)\frakr^2/\ell^2 } \, , \qquad K \ell \in \mathbb{R} \, .
\end{equation}
For $1/3\,<\,\mu/\frakr\,<\,1/2$, there are two branches of $e^\omega$ which gives rise to the same $K \ell$ when $K \ell$ is positive. They are given by
\begin{equation}\label{eqn: appendix dum2}
	e^{2\omega_\pm} \= \frac{3+\left(K^2\ell^2+9\right) \left(1-\frac{2\mu}{\frakr}\right) \pm K\ell \sqrt{\left(K^2\ell^2+9\right)\left(1-\frac{2\mu}{\frakr}\right)^2-1}}{2 \left(K^2\ell^2+9\right)\frakr^2/\ell^2 } \, , \qquad K \ell \geq 0  \, .
\end{equation}
This also means that, for $1/3\,<\, \mu/\frakr \,<\, 1/2$, there is no $e^\omega$ that leads to negative $K \ell$.

We can write the radius of the boundary as $\frakr_\text{tube} \= e^\omega \frakr$. The cosmological horizon $\frakr_{\text{c}}$ and black hole horizon $\frakr_{\text{bh}}$ can also be written in terms of $e^\omega$ and $\mu$ by
\begin{equation}
    \frakr_{\text{c}} \= \frac{2}{\sqrt{3}} \cos\left(\frac{1}{3}\cos^{-1}\left(-3\sqrt{3}e^\omega \mu\right)\right) \, , \qquad \frakr_{\text{bh}} \= \frac{2}{\sqrt{3}}\sin\left(\frac{1}{3}\sin^{-1}\left(3\sqrt{3}e^\omega \mu\right)\right) \, .
\end{equation}
Given \eqref{eqn: appendix dum1} and \eqref{eqn: appendix dum2}, one can use these formulae to investigate the behaviour of $\frakr_\text{tube}$, $\frakr_{\text{c}}$, and $\frakr_{\text{bh}}$ while keeping $K \ell$ fixed and varying $\mu/\frakr$.

\section{Details of black hole patch computations in $D=4$}  \label{black hole patch app}
In this appendix we give some more details and explicit expressions of the computations done in section \ref{sec: bh patch}, for the black hole patch solutions in $D=4$. For convenience, in this section, we will always express $\frakr_{\text{c}}$ in terms of $\frakr_{\text{bh}}$.

Regularity of the geometry near the black hole horizon determines the conformal temperature of the black hole patch to be
\begin{equation}\label{eqn: bh ds4 beta}
	\tilde{\beta} \= \frac{4\pi \frakr_{\text{bh}}\ell\sqrt{\frakr_\text{tube}\ell^2 - \frakr_\text{tube}^3 - \frakr_{\text{bh}}\ell^2 + \frakr_{\text{bh}}^3}}{\frakr_\text{tube}^{3/2} \left(\ell^2-3 \frakr_{\text{bh}}^2\right)} \, ,
\end{equation}
which is greater than zero. The conformal temperature $\tilde{\beta}^{-1}$ has a lower bound $\tilde{\beta}^{-1}_{\text{min}} \= 2\pi$, which occurs in the Nariai limit, by setting $\frakr_\text{tube} \= \ell/\sqrt{3}$ and taking $\frakr_{\text{bh}} \rightarrow \ell/\sqrt{3}$ from below. 

Below this conformal temperature, the black hole patch solution does not exist. For larger conformal temperatures, $\tilde{\beta}^{-1} \, > \, \tilde{\beta}^{-1}_{\text{min}}$, there is a one-parameter family of black hole patches. To reach the high conformal temperature regime, $\tilde{\beta}\rightarrow 0$, one can take the near horizons limit, i.e. either $\frakr_\text{tube} \rightarrow \frakr_{\text{bh}}$ or $\frakr_\text{tube} \rightarrow \frakr_{\text{c}}$, or the small black hole limit $\frakr_{\text{bh}}/\ell \rightarrow 0$.

Requiring that the boundary has a constant trace of the extrinsic curvature $K$ fixes
\begin{equation}\label{eqn: bh ds4 K}
	K \ell \= \frac{4\frakr_\text{tube}\ell^2 - 6 \frakr_\text{tube}^3-3\frakr_{\text{bh}}\ell^2 + 3 \frakr_{\text{bh}}^3}{2\frakr_\text{tube}^{3/2}\sqrt{\frakr_\text{tube}\ell^2-\frakr_\text{tube}^3-\frakr_{\text{bh}}\ell^2+\frakr_{\text{bh}}^3}} \,.
\end{equation}
There is no upper or lower bound on $K\ell$. Similarly to the pole patch, the limit of $K \ell$ approaching negative infinity corresponds to pushing the boundary to be near the cosmological horizon. For $K\ell$ going to positive infinity, the boundary is pushed near the black hole horizon.

\textbf{Black hole patch thermodynamics.} The regulated action is given by 
\begin{equation}\label{eqn: onshell bh 4d}
	I_{E\text{, reg}}^{(\text{bh})} \= - \frac{\pi \frakr_{\text{bh}} \left(4\frakr_\text{tube}\ell^2 - 3 \frakr_{\text{bh}}\ell^2 - 3 \frakr_{\text{bh}}^3\right)}{3G_N\left(\ell^2-3\frakr_{\text{bh}}^2\right)} - I_E^{(\text{pole})} \, ,
\end{equation}
where we used \eqref{eqn: bh ds4 beta} and \eqref{eqn: bh ds4 K} to simplify the expression. The corresponding conformal energy and conformal entropy for the black hole patch are given by
\begin{equation}\label{eqn: conformal energy entropy bh 4d}
		E_\text{conf} \= - \frac{\frakr_\text{tube}^{3/2}\left(2\frakr_\text{tube}\ell^2-3\frakr_{\text{bh}} \ell^2+ 3 \frakr_{\text{bh}}^3\right)}{6 G_N\ell \sqrt{\frakr_\text{tube}\ell^2-\frakr_\text{tube}^3-\frakr_{\text{bh}}\ell^2+\frakr_{\text{bh}}^3}} - \frac{I_E^{(\text{pole})}}{\tilde{\beta}} \, , \qquad\qquad
		\mathcal{S}_\text{conf} \= \frac{\pi \frakr_{\text{bh}}^2}{G_N} \, .
\end{equation}
The conformal entropy agrees with the Bekenstein-Hawking entropy $A_\text{horizon}/4G_N$ where the horizon in the formula corresponds to the black hole horizon. 

Taking a small black hole limit, we find that 
\begin{equation}
	E_\text{conf} \rightarrow \frac{\frakr_{\text{bh}}}{2G_N} \frac{\frakr_\text{tube}}{\sqrt{1-\frac{\frakr_\text{tube}^2}{\ell^2}}} \, , \qquad \qquad \text{as} \quad \frakr_{\text{bh}}/ \ell \rightarrow 0  \, .
\end{equation}
Provided that $\frakr_{\text{bh}}/2G_N$ is the physical mass of the small black hole, we can see that $E_\text{conf}$ is indeed the energy as measured by the conformal clock defined on the boundary.

The specific heat at constant $K$ of the black hole patch is given by
\begin{equation}\label{eqn: spec heat bh ds4}
	C_K = \frac{2\pi \frakr_{\text{bh}}^2 \left(-\ell^2+3 \frakr_{\text{bh}}^2\right)\left(9\frakr_{\text{bh}}^2\left(\frakr_{\text{bh}}^2-\ell^2\right)^2+16 \frakr_{\text{bh}} \left(\frakr_{\text{bh}}^2-\ell^2\right)\frakr_\text{tube}\ell^2+8\frakr_\text{tube}^2\ell^4 - 4 \frakr_\text{tube}^4\ell^2\right)}{G_N\left(\ell^2+3\frakr_{\text{bh}}^2\right)\left(9 \frakr_{\text{bh}}^2\left(\frakr_{\text{bh}}^2-\ell^2\right)^2+2\frakr_{\text{bh}}\frac{\left(-9\ell^4-10\frakr_{\text{bh}}^2\ell^2+15\frakr_{\text{bh}}^4\right)}{\left(\ell^2+3\frakr_{\text{bh}}^2\right)}\frakr_\text{tube}\ell^2+8\frakr_\text{tube}^2\ell^4-4\frakr_\text{tube}^4\ell^2\right)} \, .
\end{equation}
Note that
\begin{equation}
	C_K \,\rightarrow\, 
	\begin{cases}
		- \frac{2 \pi \frakr_{\text{bh}}^2}{G_N} \, , & \text{as} \quad \frakr_{\text{bh}}/\ell \rightarrow 0 \, , \\
		+ \frac{2 \pi \frakr_{\text{bh}}^2}{G_N} \, , & \text{as} \quad \frakr_{\text{bh}} / \frakr_\text{tube} \rightarrow 1 \, .
	\end{cases}
\end{equation}

The first limit corresponds to a small black hole and, in that case, the specific heat is negative. On the opposite limit, when the black hole size is comparable to the boundary size, then the specific heat is positive.
%
%


\section{Plots of the regulated action, $E_\text{conf}$, and $C_K$ in $D=4$}\label{sec: additional figures}

\begin{figure}[p!]
        \centering
        \subfigure[reg. action, $K\ell=-7.5$]{
                \includegraphics[height=4.5cm]{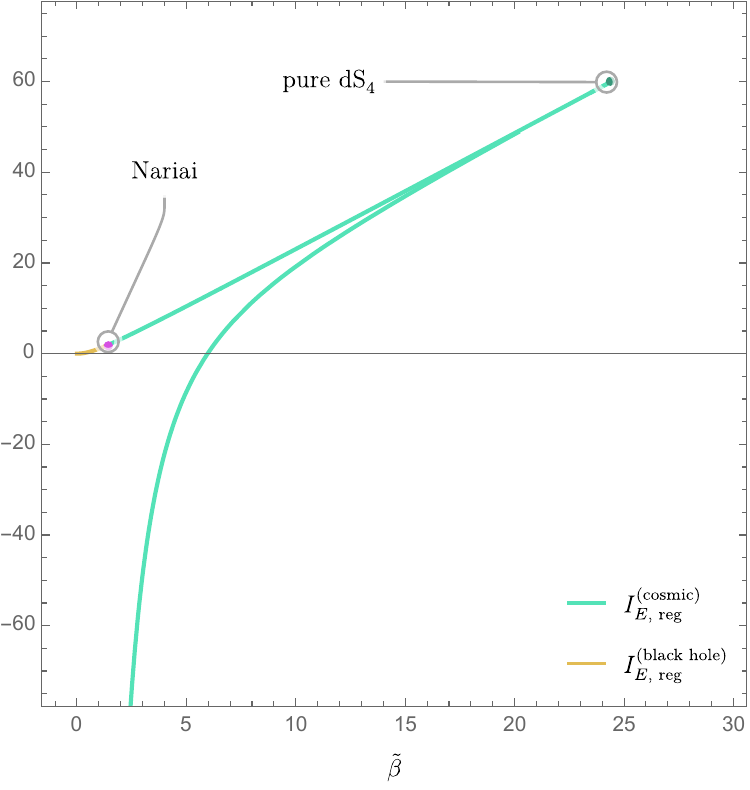}\label{fig: action -7.5}}  \quad\quad
        \subfigure[reg. action, $K\ell=-1.5$]{
                \includegraphics[height=4.5cm]{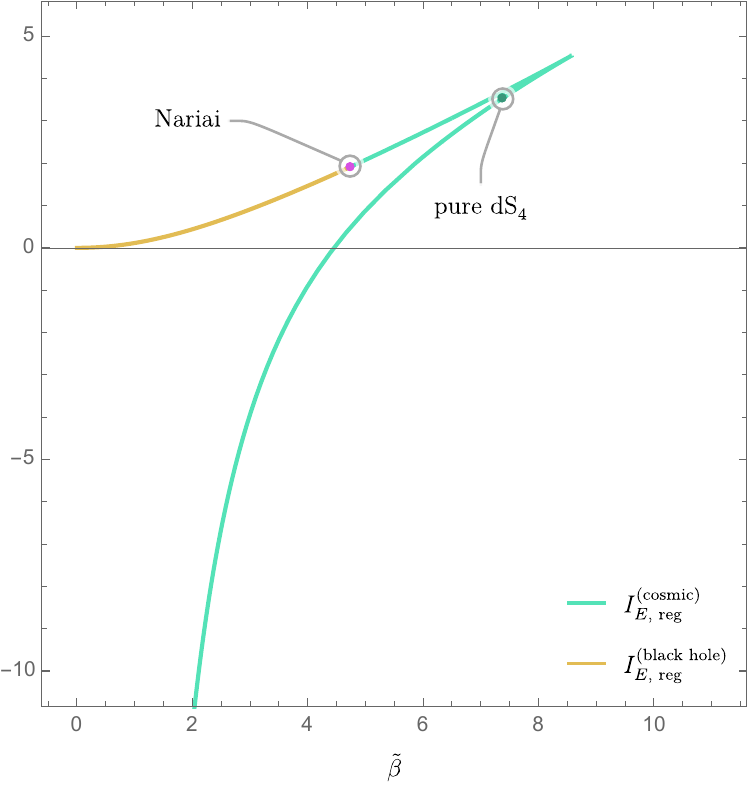}\label{fig: action -1.5}}  \quad\quad
        \subfigure[reg. action, $K\ell=6$]{
                \includegraphics[height=4.5cm]{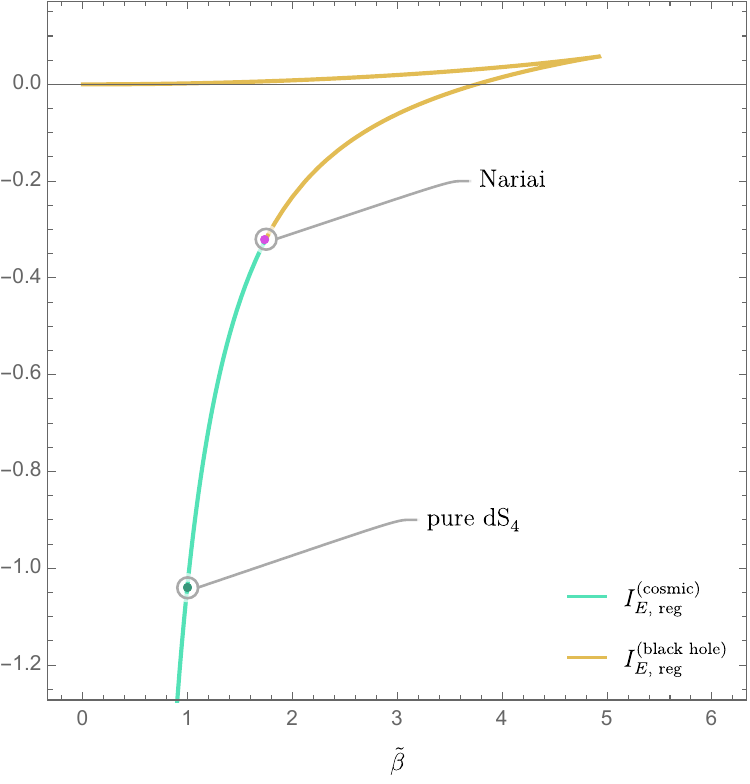}\label{fig: action 6}}  \quad\quad
        \subfigure[$E_\text{conf}$, $K\ell=-7.5$]{
                \includegraphics[height=4.5cm]{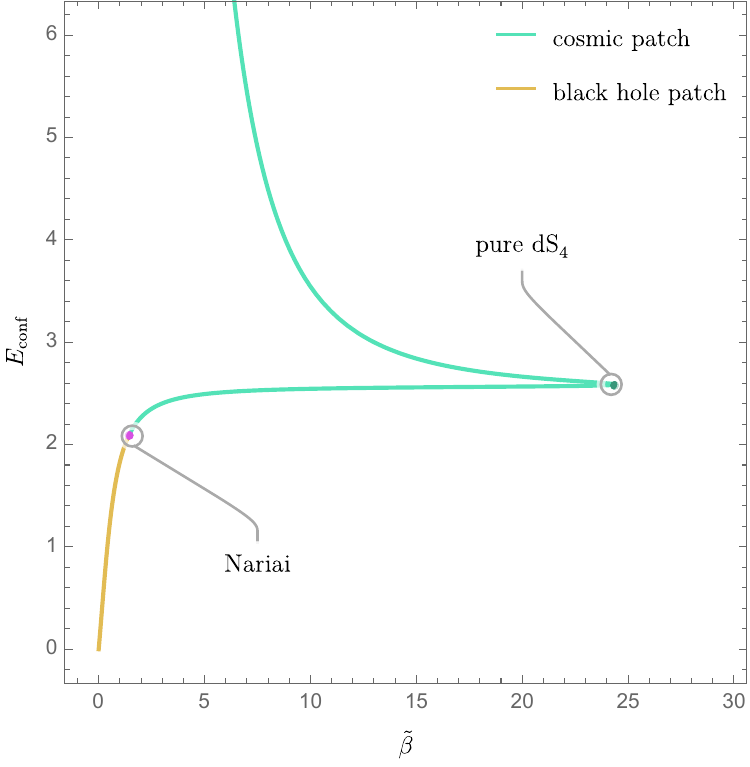}\label{fig: energy -7.5}}  \quad\quad
        \subfigure[$E_\text{conf}$, $K\ell=-1.5$]{
                \includegraphics[height=4.5cm]{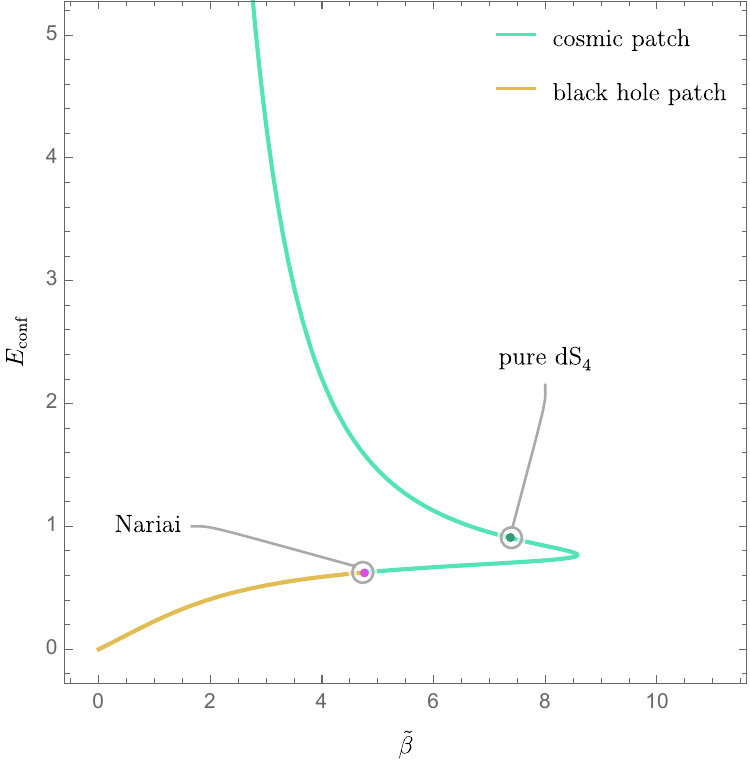}\label{fig: energy -1.5}}  \quad\quad
        \subfigure[$E_\text{conf}$, $K\ell=6$]{
                \includegraphics[height=4.5cm]{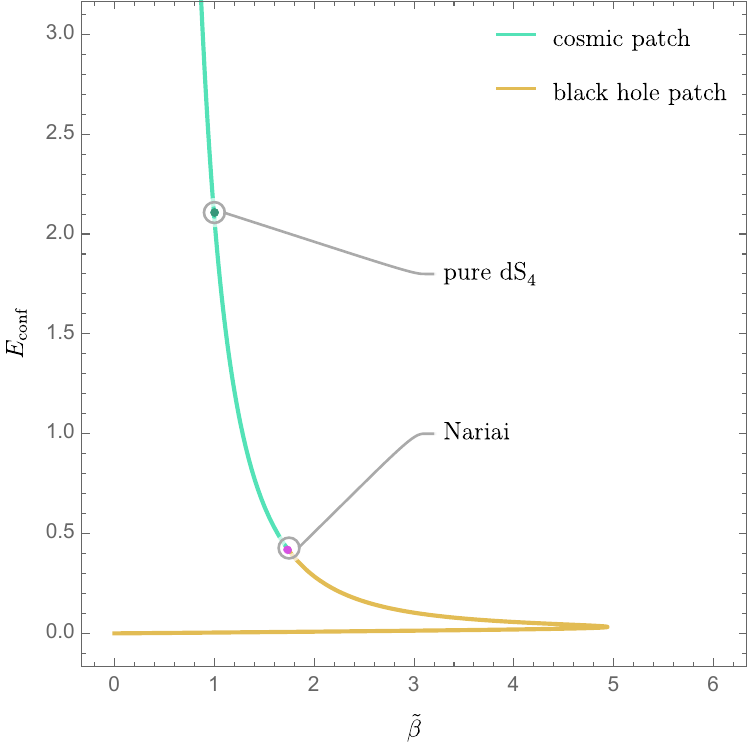}\label{fig: energy 6}}  \quad\quad
        \subfigure[$C_K$, $K\ell=-7.5$]{
                \includegraphics[height=4.5cm]{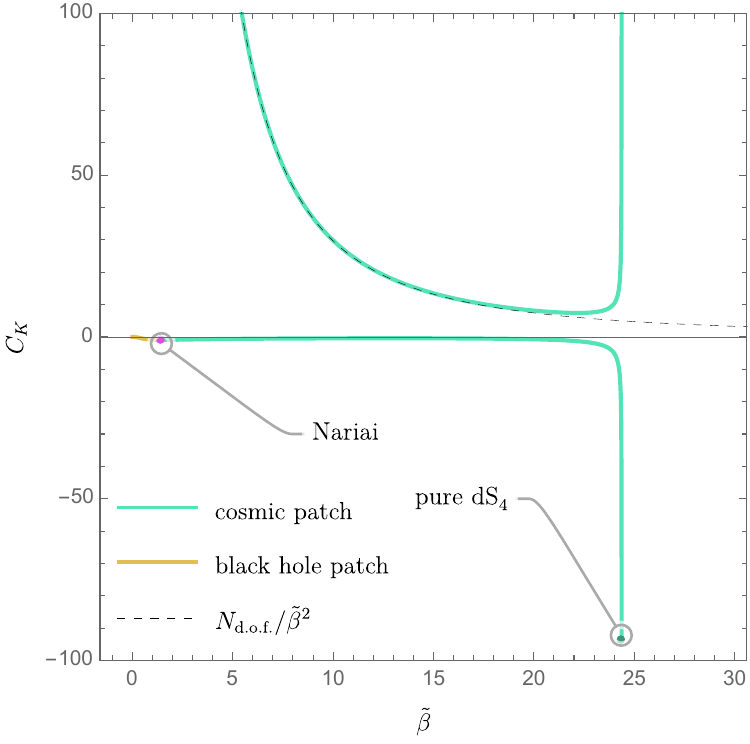}\label{fig: C -7.5}}  \quad\quad
        \subfigure[$C_K$, $K\ell=-1.5$]{
                \includegraphics[height=4.5cm]{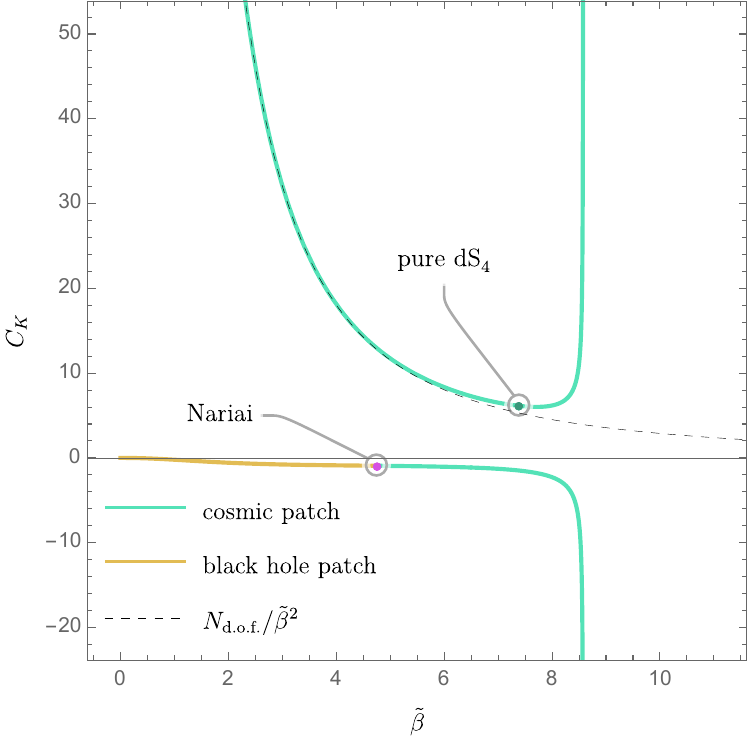}\label{fig: C -1.5}}  \quad\quad
        \subfigure[$C_K$, $K\ell=6$]{
                \includegraphics[height=4.5cm]{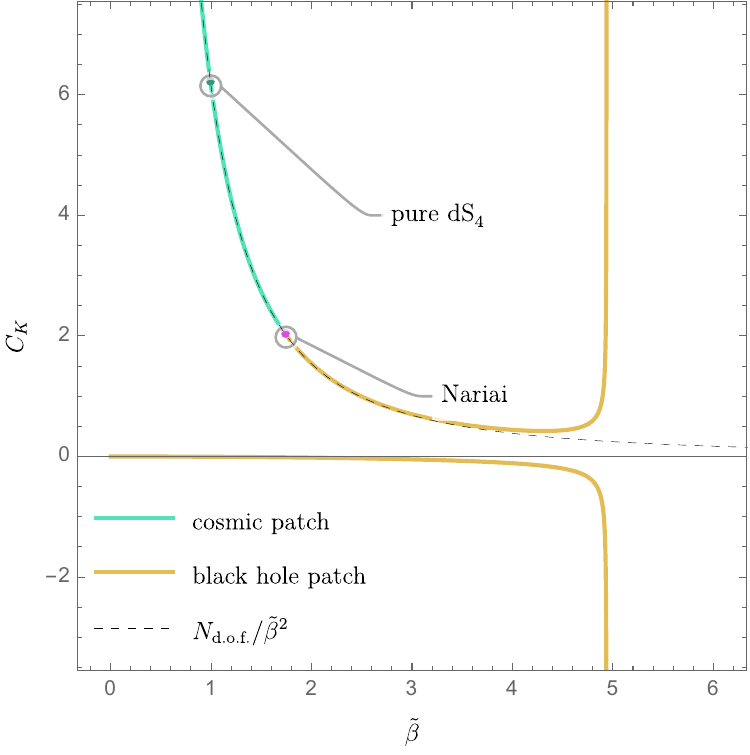}\label{fig: C 6}}      
                \caption{Plots of regulated action, $E_\text{conf}$, and $C_K$ as a function of $\tilde{\beta}$ at fixed $K\ell$. When evaluated on the cosmic (black hole) patch, the curve colour is green (yellow). The pure dS$_4$ and Nariai solutions are marked in dark green and purple dots, respectively. For the plots of $C_K$, we also display the conformal answer $N_\text{d.o.f.}/\tilde{\beta}^2$ as a black dashed curve.} \label{fig: appendix figures}
\end{figure}

In this appendix, we give numerical examples of the regulated action, conformal energy, and specific heat for dS$_4$ conformal thermodynamics as functions of $\tilde{\beta}$ for various values of $K\ell$. These are displayed in figure \ref{fig: appendix figures} for $K\ell \= -7.5$, $-1.5$, and $6.0$. These values of $K\ell$ are chosen to show pure dS$_4$ patches which are unstable, metastable, and stable, respectively. 

\section{$l\=0$ and $l\=1$ modes}\label{sec: l0/l1 modes}

In this appendix, we consider linearised dynamics of gravitational $l\=0$ and $l \=1$ modes. Following the analysis in \cite{Anninos:2023epi}, these modes are locally pure diffeomorphisms that become physical by the presence of the timelike boundary with fixed boundary data. In particular, we are interested in linearised perturbation $h_{\mu\nu}$ of the form,
\begin{equation}\label{eqn: physical diffeo perturbation}
    h_{\mu \nu} \= \nabla_\mu \xi_\nu + \nabla_\nu \xi_\mu \, ,
\end{equation}
for an arbitrary vector field $\xi^\mu$. The perturbation \eqref{eqn: physical diffeo perturbation} automatically satisfies the linearised Einstein field equation. The conditions that this perturbation preserves the conformal boundary data at $r\=\frakr$ lead to
\begin{equation}\label{eqn: bdry conds diffeo}
\begin{cases}
    \left.2\left(K_{mn} - \frac{K}{3}\bar{g}_{mn}\right)\left(\sqrt{1-\frac{r^2}{\ell^2}}\xi_r\right) + \left(\mathcal{D}_m \xi_n + \mathcal{D}_n \xi_m - \frac{2\bar{g}_{mn}}{3}\mathcal{D}^p \xi_p\right)\right|_{r=\frakr} &=\, 0 \, , \\
    \left.\left(\sqrt{1-\frac{r^2}{\ell^2}}\partial_r K - \mathcal{D}^m \mathcal{D}_m \right) \left(\sqrt{1-\frac{r^2}{\ell^2}}\xi_r\right) + \xi_m \mathcal{D}^m K \right|_{r=\frakr} &=\, 0 \, ,
\end{cases}
\end{equation}
where $K_{mn}$ and $\bar{g}_{mn}$ are the extrinsic curvature \eqref{eqn: lorentzian extrinsic} and induced metric \eqref{eqn: lorentzian induced metric} of $\Gamma$, respectively. The covariant derivative $\mathcal{D}_m$ is that associated to the induced metric $\bar{g}_{mn}$. 

At the linearised level, the perturbation \eqref{eqn: physical diffeo perturbation} is subject also to gauge redundancy in the form of the diffeomorphism
\begin{equation}\label{eqn: diffeo transf}
    x^\mu \,\to\, x^\mu + \epsilon\,\xi^{'\mu} \, , \qquad\qquad h_{\mu\nu} \,\to\, h_{\mu\nu} - \nabla_\mu \xi'_\nu - \nabla_\nu \xi'_\mu \, ,
\end{equation}
for an arbitrary vector field $\xi^{'\mu}$. Due to the presence of the boundary, the vector field $\xi^{'\mu}$ must preserve the boundary data and location of the boundary leading to \eqref{eqn: bdry conds diffeo} and $\left.\xi'^r\right|_{r=\frakr} \= 0$, respectively. This means that a large number of the perturbations \eqref{eqn: physical diffeo perturbation} obeying \eqref{eqn: bdry conds diffeo} can be gauged away by some suitable diffeomorphism \eqref{eqn: diffeo transf}. The exception is when the perturbation \eqref{eqn: physical diffeo perturbation} is constructed from the vector field $\xi^\mu$ which disturbs the location of the boundary, i.e.
\begin{equation}\label{eqn: disturbing location condition}
    \left.\xi^r\right|_{r=\frakr} \,\neq\, 0 \, .
\end{equation}
We therefore take \eqref{eqn: bdry conds diffeo} and \eqref{eqn: disturbing location condition} as boundary conditions for a physical metric perturbation \eqref{eqn: physical diffeo perturbation}.

\textbf{$l\=0$ modes.} Choosing the $h_{tr}\=0$ gauge, the general spherically symmetric ($l\=0$) vector field $\xi^\mu$ satisfying \eqref{eqn: bdry conds diffeo} and \eqref{eqn: disturbing location condition} is given by
\begin{equation}\label{eqn: l=0 sol}
    \xi_\mu dx^\mu \,\sim\, \frac{r}{\ell}\left(1-\frac{r^2}{\ell^2}\right)^{\omega^{(l=0)2}_\pm \ell^2/2}e^{-i \omega^{(l=0)}_\pm t} \left(dr - i \frac{1-r^2/\ell^2}{\omega^{(l=0)}_\pm r}\, dt\right)\, , 
\end{equation}
where the frequency $\omega^{(l=0)}_\pm \frakr \= \pm i \sqrt{2-\frac{\frakr^2}{\ell^2}}$ is purely imaginary. In the worldline limit, where $\frakr/\ell \,\to\, 0$, we match the result of $l\=0$ modes found in \cite{Anninos:2023epi}. In the strechted horizon limit, where $\frakr/\ell \,\to\, 1$, we find that these pair of modes coalesce to $\omega^{(l=0)}_\pm \ell \= \pm i$.

\textbf{$l\=1$ modes.} Taking again the $h_{tr}\=0$ gauge, the general $l\=1$ vector field $\xi^\mu$ satisfying \eqref{eqn: bdry conds diffeo} and \eqref{eqn: disturbing location condition} is given by
\begin{equation}\label{eqn: l=1 sol}
    \xi_\mu dx^\mu \,\sim\,\frac{e^{- i \omega^{(l=1)}_\pm t}}{\sqrt{1-r^2/\ell^2}} \left(\mathbb{S} \, dr + i {\omega^{(l=1)}_\pm r}\left(1-r^2/\ell^2\right) \,\mathbb{S} \, dt +\sqrt{2}\left(1-r^2/\ell^2\right) \,\mathbb{S}_i \,r \,d\Omega^i\right) \, ,
\end{equation}
where the frequency $\omega^{(l=1)}_\pm \ell \= \pm i$ is pure imaginary and $\frakr$-independent. Unlike the $l\=0$ modes, the metric perturbation constructed from \eqref{eqn: l=1 sol} is vanishing everywhere implying that \eqref{eqn: l=1 sol} is a Killing vector of the background dS$_4$. Near the worldline, where $\frakr/\ell \,\to\, 0$, these modes reproduce three translations and three Lorentz boosts of the flat spacetime. In the strectched horizon limit, where $\frakr/\ell \,\to\, 1$, we find that these modes become combinations of translations and Lorentz boosts in a local inertial frame near the boundary. 

\section{Near-horizon diffeomorphisms}\label{diffeoapp}

In this appendix, we consider metric perturbations that are locally diffomorphisms for any $l$, and we consider a boundary that is located close to the cosmological horizon.

Let us first define a near-horizon parameter $\epsilon \,\equiv\,1-\tfrac{r}{\ell}$ and a near-horizon radial coordinate $\rho\,\equiv\,\tfrac{1}{{\epsilon}} \left({1-\tfrac{r}{\ell}}\right)$. It follows that in terms of $\rho$, the boundary is located at $\rho\=1$. 

Consider the following linearised diffeomorphism with complex frequency $\omega \ell \= \pm i$,
\begin{equation}\label{eqn: supertranslation}
    \xi_\mu dx^\mu \= e^{\pm t} \mathbb{S} \,dr - e^{\pm t} \left(1-\frac{r^2}{\ell^2}\right)  \left(\pm\mathbb{S}\, dt - \partial_i \mathbb{S}\, d\Omega^i\right)\, ,
\end{equation}
where $\mathbb{S}$ here is an arbitrary angle-dependent function obeying $\left|\frac{\partial_i \mathbb{S}}{\mathbb{S}}\right|\,\ll\,\frac{1}{\epsilon}$. By considering the associated linearised metric perturbation $h_{\mu\nu} \= \nabla_\mu \xi_\nu + \nabla_\nu \xi_\mu$, we find that
\begin{equation}
    \left.\delta K(h_{\mu\nu})\right|_{r=\frakr} \= \mathcal{O}(\sqrt{\epsilon})  \, , \qquad\qquad \left.\delta h^m{}_n\right|_{r=\frakr} \= \mathcal{O}(\epsilon)\delta^m_n \, .
\end{equation}
In terms of the near-horizon coordinate, we find that
\begin{equation}
    \left.\xi^\theta\right|_{\rho=1} \= \left.\xi^\phi\right|_{\rho=1} \= \mathcal{O}(\epsilon) \, , \qquad -\frac{\ell}{2}\left.\xi^\rho\right|_{\rho=1} \= \pm \left.\xi^t\right|_{\rho=1} \= e^{\pm t/\ell} \mathbb{S} + \mathcal{O}(\epsilon)\, .
\end{equation}
This means that, in the $\epsilon\,\to\,0$ limit, the diffeomorphism \eqref{eqn: supertranslation} preserves the conformal boundary data but not the location of the boundary. In particular, in the local inertial frame, these modes become angle-dependent radial/time translations. The frequencies of these modes coincide with the complex frequency modes, $\omega \ell = \pm i$, found in the cosmological horizon limit discussed in section \ref{sec: scalar pert}.

\bibliographystyle{JHEP}
\bibliography{bibliography}

\end{document}